\documentclass[11pt]{article}

\usepackage{acl}

\usepackage{times}
\usepackage{latexsym}
\usepackage{amsmath}
\usepackage{amssymb}
\usepackage[T1]{fontenc}
\usepackage[utf8]{inputenc}
\usepackage{microtype}
\usepackage{inconsolata}
\usepackage{graphicx}
\usepackage{pifont}
\usepackage{seqsplit}
\usepackage{booktabs}
\usepackage{multirow}
\usepackage{float}
\usepackage{enumitem}
\usepackage{xcolor}
\usepackage{pifont}

\usepackage{calc}
\usepackage{makecell}
\usepackage{placeins}
\usepackage{stfloats}
\usepackage{multicol}

\usepackage{longtable}
\usepackage{listings}
\usepackage{tipa}
\usepackage{bm}

\usepackage{tcolorbox}
\tcbuselibrary{breakable}
\tcbset{
  promptbox/.style={
    breakable,
    colback=gray!10,
    colframe=gray!40,
    boxrule=0.5pt,
    arc=3pt,
    left=6pt, right=6pt, top=4pt, bottom=4pt,
    fontupper=\small\ttfamily
  }
}

\lstdefinestyle{jsonblock}{
    basicstyle=\scriptsize\ttfamily,
    backgroundcolor=\color{gray!10},
    frame=single,
    framerule=0.4pt,
    rulecolor=\color{gray!40},
    columns=flexible,
    breaklines=true,
    showstringspaces=false,
    keepspaces=true
}

\lstdefinestyle{artifactid}{
    basicstyle=\scriptsize\ttfamily,
    backgroundcolor=\color{gray!10},
    frame=single,
    framerule=0.4pt,
    rulecolor=\color{gray!40},
    columns=flexible,
    breaklines=true,
    showstringspaces=false,
    keepspaces=true
}

\newcommand{\cmark}{\textcolor{green!50!black}{\ding{51}}}
\newcommand{\pmark}{\textcolor{orange!70!black}{\textasciitilde}}
\newcommand{\xmark}{\textcolor{red!50!black}{\ding{55}}}

\usepackage{titlesec}
\titlespacing*{\section}{0pt}{6pt}{4pt}
\titlespacing*{\subsection}{0pt}{4pt}{2pt}
\titlespacing*{\subsubsection}{0pt}{4pt}{2pt}
\titlespacing*{\paragraph}{0pt}{4pt}{1em}

\title{Beyond Naturalness: Probing Automated Text-To-Speech Evaluators on Linguistically Grounded Dimensions}

\author{
  \textbf{Oluwanifemi Bamgbose\thanks{\ Corresponding author.}\thanks{\ Equal contributions}, Jash Shah$^\dagger$, Simon Rosen$^\dagger$,} \\
  \textbf{Lindsay Devon Brin, Hoang H Nguyen, Anke Koelzer, Rachel Hansen,} \\
  \textbf{Tara Bogavelli, Fanny Riols} \\
  ServiceNow \\
  \texttt{\{nifemi.bamgbose, jash.shah, simon.rosen\}@servicenow.com} \\
    \href{https://github.com/ServiceNow/BeyondNaturalnessProbe}{\texttt{Code}} \quad \href{https://servicenow.github.io/BeyondNaturalnessProbe/}{\texttt{Website}} \quad \href{https://huggingface.co/datasets/ServiceNow-AI/Beyond-Naturalness-TTS-Eval}{\texttt{Dataset}}
}

\def \audiollm{Audio-LLMs}
\def \audiollmsingular{Audio-LLM}

\def \audiollmfullsingular{Audio Large Language Model}
\def \numattributes{10}
\def \numutterances{860}

\begin{document}
\maketitle

\begin{abstract}
Automated Text-to-Speech (TTS) evaluation methods (Mean Opinion Score (MOS) predictors and \audiollmfullsingular~(\audiollmsingular) judges) are expected to reflect human perception, yet it is unclear how well they capture the distinct aspects of speech that listeners actually perceive. We deconstruct "naturalness" into a linguistically grounded annotation schema spanning \numattributes~distinct perceptual dimensions, and use it to construct the first dimension-level meta-evaluation benchmark for TTS, comprising \numutterances~utterances annotated by trained linguist raters. Results from benchmarking four MOS predictors and four \audiollmsingular~judges reveal that MOS predictors are more sensitive to acoustic signal-level properties at the prosodic and paralinguistic levels, while \audiollmsingular~judges show selective, prompt-dependent detection that favors word-level errors and does not generalize across all dimensions. Neither class reliably captures a breadth of linguistically structured, human-perceptible speech errors. Our dataset, annotation schema, and evaluation code are publicly released to support more targeted and interpretable TTS evaluation.
\end{abstract}

\section{Introduction}
Text-to-speech (TTS) systems are increasingly deployed in production contexts \cite{tan2021survey, xie2025towards}, where the naturalness of synthesized speech directly affects user experience and comprehension, with potential consequences for user engagement. Listeners are sensitive to a multidimensional set of failures; a TTS system may stress the wrong syllable in a word, break a sentence at an unnatural point, or deliver a question with the flat intonation of a statement, any of which would be immediately human-perceptible \citep{geneva23_interspeech, gutierrez2021location, lee2026speakersleuth}.

The gold standard for measuring TTS quality across these dimensions remains human perceptual judgment \cite{baughan2023mixed, ulgen2026objective}. As deployment scales, however, relying solely on human judgment becomes expensive and time consuming. Two main paradigms have emerged to automatically evaluate speech quality at scale: neural Mean Opinion Score (MOS) predictors \cite{mittag2021nisqa, saeki2022utmos}, trained to output individual, holistic quality scores that reflect human MOS; and \audiollmfullsingular~(\audiollmsingular) Judges \cite{zhang2025speechjudge, manakul2026audiojudge}, which prompt audio-capable LLMs to rate or compare audio samples. A good speech evaluator should respond to all aspects of perceptual quality to which humans attend, yet it is not clear whether models of either type do so (i.e whether individual output scores align with quality along any or all human-perceptible dimensions). For \audiollm~in particular, it additionally remains an open question whether they even have the underlying capability, and/or can be prompted, to attend to each dimension.

To better understand and identify the aspects of speech to which MOS models and \audiollm~actually attend, we construct the first meta-evaluation benchmark for automated TTS evaluators that has annotations based on linguistically-grounded, human-perceptible speech dimensions.

\begin{table}[tb]
  \centering
  \small
  \setlength{\tabcolsep}{8pt}
  \renewcommand{\arraystretch}{1.2}
  \caption{\textbf{Capability comparison with recent meta-evaluation resources.}
  \textbf{ETE}: EmergentTTS-Eval~\citep{manku2026emergenttts};
  \textbf{ITE}: InstructTTSEval~\citep{huang2025instructttseval};
  \textbf{SJ}: SpeechJudge~\citep{zhang2025speechjudge};
  \textbf{SS}: SpeakerSleuth~\citep{lee2026speakersleuth}.
  \cmark~present, \pmark~partial, \xmark~absent; partial $=$ human labels only validate a
  model judge (ETE/ITE), or controlled difficulty without injected linguistic errors (SS).
  Only ours pairs expert, aspect-level human labels with a linguistically-grounded taxonomy
  and controlled error injection, auditing both MOS predictors and Audio-LLM judges.}

  \label{tab:benchmark-comparison}
  \resizebox{\columnwidth}{!}{%
  \begin{tabular}{@{}lccccc@{}}
    \toprule
    \textbf{Capability} & \textbf{Ours} & \textbf{ETE} & \textbf{ITE} & \textbf{SJ} & \textbf{SS} \\
    \midrule
    Human-annotated ground truth                  & \cmark & \pmark & \pmark & \cmark & \pmark \\
    Dimension-level annotation per sample          & \cmark & \xmark & \xmark & \xmark & \xmark \\
    Linguistically-grounded aspect taxonomy       & \cmark & \xmark & \xmark & \xmark & \xmark \\
    Covers MOS predictors \& Audio-LLM judges     & \cmark & \xmark & \xmark & \cmark & \cmark \\
    \bottomrule
  \end{tabular}%
  }
\end{table}

Our contributions are as follows: 
\begin{itemize}[nolistsep, noitemsep, topsep=0pt, partopsep=0pt, 
leftmargin=*]
\item \textbf{Annotation Schema.} A linguistically grounded evaluation schema decomposing naturalness into \numattributes~phonologically and phonetically grounded dimensions across word, prosodic, and paralinguistic levels. This enables future evaluation to be aligned with human perception of speech quality.
\item \textbf{Meta-Evaluation Dataset.} A per-dimension annotated dataset 
of \numutterances~utterances, labeled by trained linguist raters across the \numattributes~dimensions.
\item \textbf{Dimension-Level Audit.} An evaluation of four neural MOS models as well as four~\audiollmsingular~judges across four prompting conditions, with and without reference transcripts. This reveals uneven phonetic, prosodic and paralinguistic sensitivity, with neither MOS predictors nor \audiollmsingular~judges reliably detecting errors across these dimensions under any condition tested.
\end{itemize}
Our benchmark enables researchers and practitioners to attribute TTS system failures to specific interpretable dimensions rather than opaque holistic scores. This provides a diagnostic foundation that prior evaluation frameworks have lacked.
\begin{figure*}[t]
  \centering  \includegraphics[width=\textwidth]{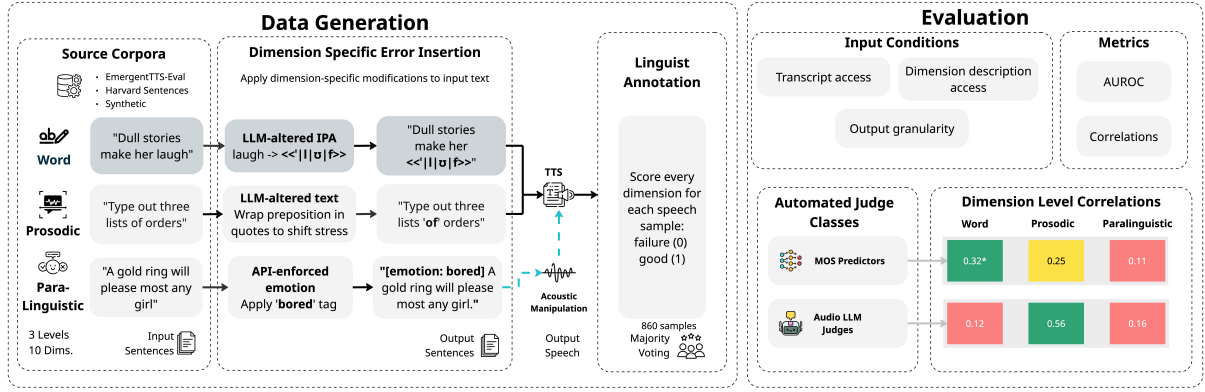}
  \caption{\textbf{Dataset Construction Pipeline.} Sentences from three sources (Harvard Sentences, EmergentTTS-Eval, synthetic text) are fed through three error-generation pathways: LLM-modified plain text, LLM-modified IPA transcriptions, and direct acoustic manipulation. All pathways pass through the TTS model to produce clean (Score 1) and error-targeted samples. Samples are then annotated by three linguists on all ten dimensions. After linguist annotation, majority filtering and downsampling produce a balanced final dataset of 860 samples.}
  \label{fig:dataset_costruction}
\end{figure*}

\section{Related Works}

\paragraph{Automated TTS Evaluation.} The dominant paradigm for scalable TTS evaluation has been MOS predictors: learning to approximate human naturalness scores from raw waveforms without a reference signal. Through the VoiceMOS Challenge series (2022–2024) \cite{huang2022voicemos, huang2024voicemos}, Semi-supervised Learning (SSL)-based MOS predictors~\citep{patton2016automos,reddy2021dnsmos,saeki2022utmos,baba2024utmosv2} have demonstrated strong system-level correlation with human judgments, approaching the practical ceiling of aggregate naturalness scoring. Alternatively, metrics such as TTSDS2~\cite{minixhofer2026ttsds2} directly compare corpus-level acoustic distributions across human and TTS-generated speech to achieve robust cross-domain TTS evaluation without fine-tuning.
However, these methods still produce holistic quality signals without grounding them in independently interpretable linguistic attributes, leaving no diagnostic information for failures. More recently, researchers have explored leveraging \audiollm~as richer judges for TTS evaluation, evaluating speaking style~\cite{chiang-etal-2025-audio}, speech quality~\cite{zhang2025speechjudge,manakul2026audiojudge,monjur2025speechqualityllm}, and speaker consistency~\cite{lee2026speakersleuth}. Yet, most implementations of LLM-based judges inherit the same limitation: scores remain holistic, and their sensitivity to the perceptual dimensions driving any given rating is unclear (\ref{tab:benchmark-comparison}).

\paragraph{TTS Evaluation Benchmarks.}
Several benchmarks have moved beyond scalar MOS towards richer TTS evaluation~\cite{manku2026emergenttts,huang2025instructttseval}, yet share a fundamental limitation: evaluation dimensions are organized around predefined scenario categories rather than a principled speech quality schema. EmergentTTS-Eval~\cite{manku2026emergenttts} introduces a model-as-a-judge benchmark spanning six pre-defined scenario types (emotions, paralinguistics, foreign words, syntactic complexity, complex pronunciation, and questions) and demonstrates that \audiollm~judges can distinguish broad scenario-level variation. InstructTTS-Eval~\cite{huang2025instructttseval} takes a complementary direction, centralizing evaluations on expressive and style-following capability of TTS systems across pre-defined dimensions such as speaking rate, pitch, and emotional expressiveness. However, neither provides human-annotated ground truth at the attribute level, making it impossible to audit whether automated evaluators track the aspects listeners actually perceive (\ref{tab:benchmark-comparison}). We bridge this gap with a human-annotated dataset grounded in 10 interpretable aspects across three linguistically-motivated quality dimensions, paired with a calibrated annotation framework, enabling direct attribution of TTS system failures to specific and interpretable aspects.

\begin{table*}[t]
\centering
\small
\setlength{\tabcolsep}{4pt}
\caption{%
  \textbf{The ten-aspect evaluation schema organized by dimension.} Full annotation
  criteria and worked examples are provided in Table ~\ref{tab:full_rubric}.
}
\label{tab:schema_intro}
\resizebox{0.9\textwidth}{!}{%
\begin{tabular}{p{2.0cm} p{3.8cm} p{8.0cm}}
\toprule
\textbf{Level} & \textbf{Dimension} & \textbf{Failure condition} \\
\midrule

\multirow{2}{2.0cm}{\textit{Word}}
  & Phonetic accuracy
  & Sound outside acceptable range of lexical target \\[3pt]
  & Lexical stress
  & Stress assigned to wrong syllable \\

\midrule

\multirow{5}{2.0cm}{\textit{Prosodic}}
  & Intonation
  & Pitch contour inappropriate to utterance type or pragmatic meaning \\[3pt]
  & Prosodic stress
  & Post-lexical prominence misplaced relative to information structure \\[3pt]
  & Prosodic boundary placement
  & Phrase boundaries inconsistent with syntactic structure \\[3pt]
  & Speech rate appropriateness
  & Tempo inappropriate to context \\

\midrule

\multirow{4}{2.0cm}{\textit{Paralinguistic}}
  & Expressiveness
  & Affective range insufficient or excessive for content \\[3pt]
  & Emotional appropriateness
  & Emotional tone inconsistent with content \\[3pt]
  & Speaker identity consistency
  & Voice characteristics vary within the utterance \\[3pt]
  & Human plausibility
  & Vocal qualities outside physically realizable range \\

\bottomrule
\end{tabular}%
}
\end{table*}

\section{Benchmark Design}
\subsection{Evaluation Schema}
\label{subsec:evaluation_schema}
Current TTS evaluation conflates perceptually distinct failure types under a single quality judgment. Terms such as ``naturalness,'' ``fluency,'' and ``clarity'' are used interchangeably in MOS elicitation, yet listeners explain their ratings by invoking factors ranging from pronunciation accuracy to affective delivery~\cite{kirkland23_ssw}, aspects that are not only perceptually distinct but defined at different levels of linguistic structure. Drawing on Crystal's distinction between linguistic and paralinguistic features~\cite{crystal1969prosodic}, we organize evaluation around three levels corresponding to where each property is linguistically specified: the lexicon, the utterance, and the speaker. We propose a ten-dimension schema in which each attribute is operationally defined and perceptually separable (see \autoref{tab:schema_intro}).

\subsubsection{Word level}
Quality at the word level is determined by whether each word contains the correct sounds and carries stress on the correct syllable. We organize this level around fixed properties of lexical entries, including both phoneme identity and stress position. Errors at this level (whether phonetic or stress-related) constitute word identity errors regardless of sentence context~\cite{jesse2017english}.

\textbf{Phonetic accuracy:} Whether the sounds in each word fall within the acceptable range of realization for their lexical targets.

\textbf{Lexical stress:} Whether primary stress falls on the correct syllable of each word, including appropriate vowel reduction in unstressed syllables.

\subsubsection{Prosodic level}
Quality at the prosodic level is determined by whether suprasegmental properties are appropriate for the syntactic and semantic properties of the utterance. Prosody is a perceptually separable dimension that listeners isolate from overall speech quality~\cite{gutierrez2021location} and one where current systems remain measurably below natural speech~\cite{vallesperez21_interspeech}. 

We define the dimensions below based on linguistic function.

\textbf{Intonation:} The pitch contour over phrases and sentences, encoding utterance type (e.g. question, statement) and discourse structure (e.g. topic boundaries, turn-taking signals).

\textbf{Prosodic stress:} Prominence on specific words or phrases relative to the rest of the utterance, determined by information structure. 

\textbf{Prosodic boundary appropriateness:} The chunking of the utterance into prosodic phrases aligned with syntactic structure, realized through boundary tones, lengthening, and pause.

\textbf{Speech rate appropriateness:} The overall pace of delivery, which can fail without any corresponding failure in pitch or prosodic structure.

\subsubsection{Paralinguistic level}
Quality at the paralinguistic level is determined by whether the speech conveys speaker characteristics appropriate to the intended speaker and content.
Although paralinguistic dimensions share audio properties with prosody, such as pitch and timing, they convey information about the speaker rather than the linguistic message. Following~\citet{crystal1969prosodic}, we distinguish speaker-stable voice quality features from affective features that modulate in response to content and context (see Appendix~\ref{app:paralinguistic-framework}). Where no external reference speaker exists, dimensions are assessed relative to the system's own output.

\textbf{Emotional appropriateness:} The degree to which the emotional tone matches the affective content of the utterance.

\textbf{Expressiveness:} Natural variation in delivery style and vocal energy, assessed relative to the system itself.

\textbf{Human plausibility:} The degree to which vocal qualities fall within the range of physically realizable human production.

\textbf{Speaker identity consistency:} Stability of perceived vocal characteristics across the utterance, assessed relative to the system itself.

\begin{table*}[t]
  \centering
  \small
  \setlength{\tabcolsep}{5pt}
  \caption{\textbf{Error generation method, sample counts, and inter-annotator agreement per dimension.}
    $n$ is the balanced evaluation set after majority-class downsampling to a 50/50
    positive/negative split (positives $=$ negatives $= n/2$ per dimension).
    Krippendorff's $\alpha$ is reported on this set with 95\% bootstrap CIs (bold: $\alpha$<0.0667). Total: $n = 860$.}
  \label{tab:generation}
  \resizebox{\textwidth}{!}{%
  \begin{tabular}{llllp{4.6cm}rl}
    \toprule
    \textbf{Level} & \textbf{Dimension} & \textbf{Method} & \textbf{Text Source} & \textbf{Manipulation}
      & \bm{$n$} & \textbf{Krippendorff's} \bm{$\alpha$} \\
    \midrule
    \multirow{2}{*}{Word}
     & Phonetic Accuracy & LLM-altered IPA & Harvard Sentences & Phoneme substitution & 110 & \textbf{0.767} \textbf{\tiny{[0.660, 0.852]}} \\[3pt]
     & Lexical Stress & LLM-altered IPA & Harvard Sentences & Stress mark transposition & 90 & \textbf{0.821} \textbf{\tiny{[0.728, 0.907]}} \\
    \midrule
    \multirow{4}{*}{Prosodic}
     & Intonation & Acoustic manipulation & EmergentTTS-Eval & F0 contour manipulation & 80 & 0.469 \tiny{[0.332, 0.611]} \\[3pt]
     & Prosodic Stress & LLM-altered plain text & Harvard Sentences & Function word quoting & 102 & 0.658 \tiny{[0.542, 0.766]} \\[3pt]
     & Prosodic Boundary Placement & LLM-altered plain text & Synthetically generated & Structurally targeted sentences & 88 & \textbf{0.727} \textbf{\tiny{[0.604, 0.834]}} \\[3pt]
     & Speech Rate Appropriateness & Acoustic manipulation & Harvard Sentences & Duration scaling & 128 & \textbf{0.707} \textbf{\tiny{[0.605, 0.801]}} \\
    \midrule
    \multirow{4}{*}{Paralinguistic}
     & Emotional Appropriateness & API-enforced emotion & Synthetically generated & Text–emotion mismatch via Cartesia tag & 46 & 0.596 \tiny{[0.400, 0.767]} \\[3pt]
     & Expressiveness & Acoustic manipulation & Harvard Sentences & F0 manipulation, loudness compression, vowel shortening & 94 & 0.460 \tiny{[0.332, 0.597]} \\[3pt]
     & Speaker Identity Consistency & Acoustic manipulation & Harvard Sentences & Cross-speaker fading & 66 & \textbf{0.713} \textbf{\tiny{[0.577, 0.836]}} \\[3pt]
     & Human Plausibility & Acoustic manipulation & Harvard Sentences & Inserted audio distortions and glitches & 56 & \textbf{0.710} \textbf{\tiny{[0.565, 0.849]}} \\
    \midrule
    & \textbf{Total} & & & & \textbf{860} & \\
    \bottomrule
  \end{tabular}%
    }
\end{table*}

\subsection{Dataset Construction }
Our dataset consists of 860 samples of annotated and majority-voted audios. We generated synthetic speech using Cartesia Sonic-3 and ElevenLabs Multilingual v2, and introduced controlled errors through text modification and acoustic manipulation. We note that this design prioritizes experimental control over resemblance to spontaneous TTS errors, aimed at yielding an upper-bound estimate of detection sensitivity under known error conditions. Below we describe our source corpora, annotation scale, error generation methods, and human annotation procedure.

\paragraph{Source corpora.}
To systematically cover the annotation schema dimensions, we selected utterances from two existing corpora and synthetically generated additional texts where needed (Table~\ref{tab:generation}). Harvard Sentences~\cite{ieee1969harvard} provide 720 phonetically balanced sentences covering the English phoneme inventory broadly, well-suited to word-level and some prosodic dimensions. EmergentTTS-Eval~\cite{manku2026emergenttts} provides sentences with complex syntactic structures and varied question types (useful for prosodic boundary and intonation dimensions) alongside sentences with explicitly defined affective content, used for emotional appropriateness. For dimensions lacking appropriate utterances in these sources, we synthetically generated sample texts (Appendix \ref{app:synthetic_text}). 

\paragraph{Scale design.}
All dimensions are rated as \textbf{binary (0/1)}, where 0 means any error is present and 1 means the audio is perfect, following the annotation schema.

\paragraph{Error generation pipeline.}
\label{sec:error-gen-pipeline}
For each source sentence, we generated utterances at the intended quality level by introducing targeted errors via one of three pathways (see Figure~\ref{fig:dataset_costruction}):

\begin{itemize}[nolistsep, noitemsep, topsep=0pt, partopsep=0pt, 
leftmargin=*]
    \item \textbf{LLM-altered IPA.} Used for Phonetic Accuracy and Lexical Stress. An LLM (GPT-5) produces a modified IPA transcription instantiating the intended error, passed to Cartesia Sonic-3~\cite{cartesia2025sonic3} for IPA-driven synthesis (e.g., phoneme substitution: \textit{seed}~\mbox{\textipa{/sid/} \textrightarrow{} \textipa{/sib/}}; stress transposition: \textit{harvest}~\mbox{\textipa{/"hAr.vIst/} \textrightarrow{} \textipa{/hAr."vIst/}}).

  \item \textbf{LLM-altered plain text.} Used for Prosodic Stress and Prosodic Boundary Placement. For stress, quotation marks around LLM-selected function words induce atypical prominence (e.g., \textit{Type out three lists ``of'' orders}); for boundaries, commas split LLM-selected syntactic constituents (e.g., \textit{I dropped my, keys}).

    \item \textbf{API-enforced emotion.} Used for Emotional Appropriateness. Sentences with a predetermined emotion are synthesized via Cartesia with deliberately mismatched emotion tags (e.g., \textit{My beloved dog passed away quietly in his sleep tonight.}\ with tag \textit{euphoric}).

    \item \textbf{Acoustic Manipulation.} Used for Intonation, Speech Rate, Expressiveness, Speaker Identity Consistency, and Human Plausibility. Praat-based manipulation~\cite{boersma2001praat} via Parselmouth~\cite{jadoul2018introducing} introduces targeted perturbations to a clean synthesis baseline: F0 contour manipulation for intonation; duration scaling for speech rate; combined F0, loudness, and vowel duration modification for expressiveness; cross-speaker fading for speaker identity; and inserted distortions for human plausibility. We treat these as controlled upper-bound cases: a judge that cannot detect a deliberately introduced perturbation cannot detect subtler naturally occurring deviations of the same type.
\end{itemize}

\noindent The prompts and details of the acosutic modification are provided in Appendix~\ref{app:llm_prompts} and Appendix~\ref{app:ipa_tts} respectively.

\subsection{Human Annotation}
\label{sec:annotation}

\paragraph{Raters.}
Each sample was annotated by three professional linguists with experience in analytical work with natural language data. Three raters is the practical minimum for stable Krippendorff's $\alpha$ estimation; sample counts per dimension are reported in Table~\ref{tab:generation}.

\paragraph{Pilot and guideline refinement.}
To reduce guideline ambiguity~\cite{10.1613/jair.1.12752}, we conducted a pilot annotation ($n=50$ samples) with all raters and iteratively refined the guidelines after collective review of disagreements.

\paragraph{Procedure.}
Raters are initially presented with the audio only, but have the option to toggle the transcript. Every sample is rated on all applicable dimensions, not only the dimension it was generated to target. This yields cross-dimensional contamination data and allows us to quantify error co-occurrence and confirm the lack of confounds (see Appendix Section \ref{app:acoustic-overlap} and Figure~\ref{fig:correlation_by_dimension}). 

\paragraph{Inter-annotator agreement.}
Table~\ref{tab:generation} reports Krippendorff's $\alpha$~\cite{krippendorff2011computing} on the construct-aligned subset ($n = 46$-$128$ per dimension), after class balancing, with 95\% bootstrap confidence intervals (1117 resamples, percentile method).

\subsection{Dataset Finalization}

\paragraph{Aggregation and downsampling.}
Ground truth annotations for each sample, for each dimension, were determined by majority vote. Each intended (analysis) dimension was downsampled to a balanced dataset of positive and negative samples, randomly sampling from the larger class with priority for consensus annotations. 
Final dataset statistics, including per-dimension sample counts, are reported in Table~\ref{tab:generation}.

\section{Experiments}

\subsection{Judges Evaluated}

We evaluate two classes of automated judge that differ in output type 
and evaluation mechanism.

\paragraph{MOS predictors.}
We include UTMOSv2~\cite{baba2024utmosv2}, DNSMOS-Pro~\cite{cumlin24_interspeech}, NISQA ~\cite{mittag2021nisqa},
and Audiobox-Aesthetics~\cite{tjandra2025meta}. Each produces either a 
single continuous scalar per audio or dimension-wise scores.

\paragraph{\audiollmsingular~judges.}
We evaluate a set of open- and closed-source audio large language 
models: Gemini 3.5 Flash , Gemini 3 Flash,
Qwen3-Omni-30B-A3B and Step-Audio-2-Mini. \audiollmsingular~ 
judges are prompt-sensitive, motivating the systematic prompting 
conditions described in 
Section~\ref{sec:prompting}.

\subsection{Prompting Strategy Conditions}
\label{sec:prompting}

We define four prompting conditions varying in the amount of schema 
guidance provided and the granularity of the expected output. In all 
conditions, models produce binary labels matching the ground-truth 
annotation format. Each judge is queried three times per sample per 
condition, with the majority label taken as the final prediction. All 
conditions are run with and without the reference transcript.

\paragraph{Condition 1: Underspecified MOS-style prompt.}
The model is prompted to rate the "naturalness" of the sample without a detailed schema, producing a single holistic score analogous to a neural MOS predictor. This condition establishes what the judge attends to in the absence of detailed task guidance, and serves as the primary baseline.

\paragraph{Condition 2a: Schema-guided prompt, single score.}
The model is prompted with the full 
\numattributes-dimension schema but instructed to produce a single overall score. Since this yields one score per sample, we correlate it against human labels independently for each dimension, treating the same predicted score as the candidate signal for each. This condition reveals whether schema exposure alone is sufficient to direct the judge toward the right perceptual dimensions.

\paragraph{Condition 2b: Schema-guided prompt, per-dimension scores.}
The model is prompted with the full schema and instructed to produce a score for each dimension, enabling direct per-dimension correlation against ground-truth labels. This condition establishes whether models can score all dimensions simultaneously when given the a detailed schema.

\paragraph{Condition 3: Isolated per-dimension prompts.}
The model is prompted for each dimension individually. This condition reveals which dimensions the judge can reliably detect without interference from the full schema. It tests dimension-specific sensitivity in its purest form.

Full prompt text for all conditions is provided in Appendix~\ref{sec:prompt_details}.

\begin{table*}[htbp]
  \centering
  \footnotesize
  \renewcommand{\arraystretch}{0.85}
  \setlength{\tabcolsep}{3pt}
  \caption{\textbf{Per-dimension Kendall's $\tau$ against human ground truth per evaluation dimension.} \textsc{\audiollm} judge rows use the C1 naturalness prompt (single holistic score, closest MOS analogue). Full table with 95\% bootstrap CI is provided in Table \ref{tab:tau_bridge_ci_inline} (Appendix).}
  \label{tab:tau_bridge}
  \resizebox{\textwidth}{!}{%
  \begin{tabular}{l | ll | llll | llll}
    \toprule
    \textbf{Model} & \multicolumn{2}{c}{\textbf{Word}} & \multicolumn{4}{c}{\textbf{Prosodic}} & \multicolumn{4}{c}{\textbf{Paralinguistic}} \\
    \cmidrule(lr){2-3} \cmidrule(lr){4-7} \cmidrule(lr){8-11}
     & \textbf{Phon.} & \textbf{Lex.} & \textbf{Inton.}$^\S$ & \textbf{Pros.S}$^\S$ & \textbf{Pros.B} & \textbf{Rate} & \textbf{Emo.}$^\S$ & \textbf{Expr.}$^\S$ & \textbf{Spk.} & \textbf{Hum.} \\
    \midrule
    \midrule
    \multicolumn{11}{c}{MOS Models} \\
    \midrule
    AudioBox$_{\mathrm{CE}}$ & -0.051 & \phantom{-}0.161 & \phantom{-}\textbf{0.371}{\scriptsize$^{***}$} & \textbf{-0.241}{\scriptsize$^{**}$} & -0.036 & \phantom{-}\textbf{0.550}{\scriptsize$^{***}$} & \phantom{-}0.050 & \phantom{-}\textbf{0.187}{\scriptsize$^{*}$} & \textbf{-0.270}{\scriptsize$^{**}$} & \phantom{-}\textbf{0.606}{\scriptsize$^{***}$} \\
    AudioBox$_{\mathrm{CU}}$ & -0.052 & \phantom{-}0.015 & \phantom{-}0.098 & \phantom{-}0.068 & \phantom{-}0.048 & \phantom{-}\textbf{0.485}{\scriptsize$^{***}$} & -0.025 & \phantom{-}\textbf{0.418}{\scriptsize$^{***}$} & -0.177 & \phantom{-}\textbf{0.511}{\scriptsize$^{***}$} \\
    AudioBox$_{\mathrm{PC}}$ & -0.058 & -0.019 & -0.099 & \phantom{-}\textbf{0.337}{\scriptsize$^{***}$} & \phantom{-}0.152 & \phantom{-}0.020 & -0.050 & \phantom{-}\textbf{0.326}{\scriptsize$^{***}$} & -0.142 & \textbf{-0.615}{\scriptsize$^{***}$} \\
    AudioBox$_{\mathrm{PQ}}$ & -0.088 & \phantom{-}0.038 & \phantom{-}\textbf{0.272}{\scriptsize$^{**}$} & \textbf{-0.169}{\scriptsize$^{*}$} & -0.079 & \phantom{-}\textbf{0.447}{\scriptsize$^{***}$} & \phantom{-}0.015 & \phantom{-}\textbf{0.304}{\scriptsize$^{***}$} & \textbf{-0.379}{\scriptsize$^{***}$} & \phantom{-}\textbf{0.606}{\scriptsize$^{***}$} \\
    DNSMOS-Pro$_{\mathrm{BVCC}}$ & -0.112 & -0.055 & \phantom{-}0.092 & \phantom{-}0.121 & \phantom{-}0.179 & \textbf{-0.147}{\scriptsize$^{*}$} & -0.061 & -0.072 & -0.196 & \phantom{-}\textbf{0.450}{\scriptsize$^{***}$} \\
    DNSMOS-Pro$_{\mathrm{VCC}}$ & \phantom{-}0.005 & -0.121 & \phantom{-}0.026 & -0.048 & -0.038 & \phantom{-}0.071 & \phantom{-}0.128 & \phantom{-}\textbf{0.406}{\scriptsize$^{***}$} & -0.092 & \phantom{-}\textbf{0.397}{\scriptsize$^{***}$} \\
    NISQA & \phantom{-}0.027 & -0.047 & \phantom{-}0.068 & \phantom{-}0.142 & \phantom{-}0.099 & \phantom{-}\textbf{0.236}{\scriptsize$^{**}$} & -0.015 & \phantom{-}\textbf{0.347}{\scriptsize$^{***}$} & -0.036 & \phantom{-}\textbf{0.377}{\scriptsize$^{***}$} \\
    UTMOSv2 & \phantom{-}0.000 & \phantom{-}0.087 & \phantom{-}\textbf{0.289}{\scriptsize$^{**}$} & \phantom{-}\textbf{0.200}{\scriptsize$^{*}$} & -0.024 & \phantom{-}\textbf{0.390}{\scriptsize$^{***}$} & -0.062 & \phantom{-}0.099 & \phantom{-}0.050 & \phantom{-}\textbf{0.313}{\scriptsize$^{**}$} \\
    \midrule
    \multicolumn{11}{c}{AudioLLM Models} \\
    \midrule
    Gemini 3 Flash$^\dag$ & \phantom{-}\textbf{0.323}{\scriptsize$^{***}$} & \phantom{-}\textbf{0.251}{\scriptsize$^{*}$} & \phantom{-}0.125 & \phantom{-}0.125 & -0.081 & \phantom{-}0.123 & \phantom{-}0.089 & \phantom{-}\textbf{0.274}{\scriptsize$^{**}$} & \phantom{-}0.036 & \phantom{-}0.202 \\
    Gemini 3 Flash$^\triangle$ & -0.022 & \phantom{-}0.089 & \textbf{-0.225}{\scriptsize$^{*}$} & \phantom{-}0.099 & -0.104 & \phantom{-}0.091 & \phantom{-}0.000 & -0.050 & \phantom{-}0.036 & \phantom{-}0.121 \\
    Gemini 3.5 Flash$^\dag$ & \phantom{-}\textbf{0.412}{\scriptsize$^{***}$} & \phantom{-}\textbf{0.312}{\scriptsize$^{**}$} & \phantom{-}0.083 & \phantom{-}\textbf{0.227}{\scriptsize$^{*}$} & \phantom{-}\textbf{0.276}{\scriptsize$^{*}$} & \phantom{-}0.116 & \phantom{-}0.255 & \phantom{-}0.170 & \phantom{-}\textbf{0.254}{\scriptsize$^{*}$} & \phantom{-}0.238 \\
    Gemini 3.5 Flash$^\triangle$ & -0.091 & \phantom{-}0.082 & -0.026 & \phantom{-}0.111 & \phantom{-}0.230 & \phantom{-}0.000 & \phantom{-}0.045 & \phantom{-}0.135 & \phantom{-}0.068 & \phantom{-}0.219 \\
    Qwen3 Omni$^\dag$ & \phantom{-}0.000 & -0.124 & -0.113 & \phantom{-}0.039 & \phantom{-}0.064 & \phantom{-}\textbf{0.214}{\scriptsize$^{*}$} & \phantom{-}0.105 & \phantom{-}0.061 & \phantom{-}\textbf{0.293}{\scriptsize$^{*}$} & \phantom{-}0.000 \\
    Qwen3 Omni$^\triangle$ & \phantom{-}0.173 & \phantom{-}0.192 & \phantom{-}0.026 & -0.103 & \phantom{-}0.262 & \phantom{-}\textbf{0.300}{\scriptsize$^{***}$} & \phantom{-}0.088 & \phantom{-}0.138 & \phantom{-}0.228 & \phantom{-}0.237 \\
    Step Audio 2 Mini$^\dag$ & -0.018 & \phantom{-}0.022 & -0.128 & -0.080 & \phantom{-}0.116 & -0.096 & -0.132 & \phantom{-}0.065 & \phantom{-}0.129 & \phantom{-}0.073 \\
    Step Audio 2 Mini$^\triangle$ & \phantom{-}0.019 & \phantom{-}0.023 & \phantom{-}0.100 & \phantom{-}0.040 & -0.111 & -0.078 & -0.039 & -0.021 & \phantom{-}0.095 & \phantom{-}0.036 \\
    \bottomrule
  \end{tabular}%
  }

  \vspace{4pt}
  \parbox{\textwidth}{\footnotesize
    \textbf{Model Subscripts:}: CE\,=\,Content Enjoyment; CU\,=\,Content Usefulness; PC\,=\,Production Complexity; PQ\,=\,Production Quality. DNSMOS-Pro subscripts: BVCC\,=\,trained on BVCC corpus; VCC\,=\,trained on VCC2018 corpus.
    \textsc{AudioLLM} rows use the C1 naturalness prompt.
    \textbf{Columns:} Phon.\,=\,Phonetic Accuracy; Lex.\,=\,Lexical Stress; Inton.\,=\,Intonation; Pros.S\,=\,Prosodic Stress; Pros.B\,=\,Prosodic Boundary; Rate\,=\,Speech Rate; Emo.\,=\,Emotional Appropriateness; Expr.\,=\,Expressiveness; Spk.\,=\,Speaker Identity; Hum.\,=\,Human Plausibility.
    \textbf{Transcript:}
    $^\dag$\,=\,with transcript; $^\triangle$\,=\,without transcript.
    \textbf{Bold} = significant ($p < .05$). $^{*}p<.05$, $^{**}p<.01$, $^{***}p<.001$ (two-sided).
    \textbf{Inter-annotator agreement:} $^\S$ = Krippendorff's $\alpha$<0.667 (see Table 3).}
\end{table*}

\subsection{Metrics}
For each (rating dimension $\times$ model) pair, we evaluate how well model scores track binary human ground-truth labels.

\paragraph{Per-dimension effect size and significance.} Kendall's $\tau_b$ serves as a unified comparison metric for per-dimension effect sizes across both model types: for continuous models (MOS predictors) it measures rank-order correlation against binary labels, and for binary models (\audiollm) it is algebraically equivalent to Pearson's $\phi$. This equivalence enables consistent comparison of continuous and binary models within the same evaluation framework.

To test significance, for MOS predictors (continuous output), we report \textbf{Mann-Whitney $U$}, which tests for distributional separation between score distributions of positive and negative labels. For \audiollm~(binary output), we also report \textbf{McNemar's test}, which tests for marginal symmetry between model predictions and ground truth.

For MOS predictors, we additionally report AUROC, which is more naturally interpretable for continuous output and is equivalent to a normalized Mann-Whitney U statistic. AUROC rankings are consistent with effect sizes reported by Kendall's $\tau_b$ across predictors (Spearman's $\rho$ = 0.995).

\noindent 

\subsection{Main Findings}

\begin{table*}[t]
  \centering
  \footnotesize
  \renewcommand{\arraystretch}{0.85}
  \setlength{\tabcolsep}{3pt}
  \caption{\textbf{Per-dimension Kendall's $\tau$ for \textsc{\audiollmsingular} judges using C2a, C2b, and C3 prompts} (with and without a provided schema, single or per-dimension outputs). Undefined $\tau$ (model predicted same value for all samples) indicated by (.). Full table with 95\% bootstrap CI is provided in Table \ref{tab:tau_lalm_ci_inline} (Appendix).}
  \label{tab:tau_lalm_c2c3}
  \resizebox{\textwidth}{!}{%
  \begin{tabular}{l | ll | llll | llll}
    \toprule
    \textbf{Model} & \multicolumn{2}{c}{\textbf{Word}} & \multicolumn{4}{c}{\textbf{Prosodic}} & \multicolumn{4}{c}{\textbf{Paralinguistic}} \\
    \cmidrule(lr){2-3} \cmidrule(lr){4-7} \cmidrule(lr){8-11}
     & \textbf{Phon.} & \textbf{Lex.} & \textbf{Inton.}$^\S$ & \textbf{Pros.S}$^\S$ & \textbf{Pros.B} & \textbf{Rate} & \textbf{Emo.}$^\S$ & \textbf{Expr.}$^\S$ & \textbf{Spk.} & \textbf{Hum.} \\
    \midrule
    \midrule
    \multicolumn{11}{c}{C2a — Multi-dimension schema; single holistic score} \\
    \midrule
    Gemini 3 Flash$^\dag$ & \phantom{-}\textbf{0.292}{\scriptsize$^{**}$} & \phantom{-}0.037 & \phantom{-}0.197 & \phantom{-}0.141 & \phantom{-}0.064 & \phantom{-}0.000 & \phantom{-}0.264 & \phantom{-}0.142 & \phantom{-}0.177 & \phantom{-}0.135 \\
    Gemini 3 Flash$^\triangle$ & \phantom{-}\textbf{0.206}{\scriptsize$^{*}$} & \phantom{-}\textbf{0.327}{\scriptsize$^{**}$} & -0.160 & \phantom{-}0.000 & \phantom{-}\textbf{0.283}{\scriptsize$^{*}$} & \phantom{-}0.172 & \phantom{-}0.129 & \phantom{-}\textbf{0.211}{\scriptsize$^{*}$} & \phantom{-}0.124 & \phantom{-}0.000 \\
    Gemini 3.5 Flash$^\dag$ & \phantom{-}\textbf{0.486}{\scriptsize$^{***}$} & \phantom{-}\textbf{0.294}{\scriptsize$^{**}$} & \phantom{-}0.119 & \phantom{-}\textbf{0.216}{\scriptsize$^{*}$} & \phantom{-}0.064 & \phantom{-}0.000 & \phantom{-}\textbf{0.396}{\scriptsize$^{**}$} & -0.031 & \phantom{-}0.122 & \phantom{-}0.146 \\
    Gemini 3.5 Flash$^\triangle$ & \phantom{-}\textbf{0.396}{\scriptsize$^{***}$} & \phantom{-}\textbf{0.438}{\scriptsize$^{***}$} & -0.052 & \phantom{-}0.064 & -0.104 & \phantom{-}0.102 & \phantom{-}0.099 & \phantom{-}0.170 & \phantom{-}\textbf{0.296}{\scriptsize$^{*}$} & \phantom{-}0.115 \\
    Qwen3 Omni$^\dag$ & \phantom{-}0.097 & \phantom{-}0.146 & \phantom{-}0.052 & -0.101 & \phantom{-}0.192 & . & . & \phantom{-}0.182 & \phantom{-}\textbf{0.435}{\scriptsize$^{***}$} & \phantom{-}0.000 \\
    Qwen3 Omni$^\triangle$ & \phantom{-}\textbf{0.194}{\scriptsize$^{*}$} & \phantom{-}\textbf{0.212}{\scriptsize$^{*}$} & . & -0.141 & \phantom{-}0.000 & . & . & . & -0.124 & \phantom{-}0.135 \\
    Step Audio 2 Mini$^\dag$ & \phantom{-}0.000 & \phantom{-}0.186 & . & -0.101 & \phantom{-}0.000 & \phantom{-}0.000 & \phantom{-}0.213 & -0.147 & \phantom{-}\textbf{0.413}{\scriptsize$^{***}$} & \phantom{-}0.135 \\
    Step Audio 2 Mini$^\triangle$ & \phantom{-}0.041 & \phantom{-}0.000 & \phantom{-}0.160 & -0.039 & -0.111 & \textbf{-0.221}{\scriptsize$^{*}$} & \phantom{-}0.000 & -0.165 & \phantom{-}0.061 & -0.153 \\
    \midrule
    \multicolumn{11}{c}{C2b — Multi-dimension schema; one score per dimension} \\
    \midrule
    Gemini 3 Flash$^\dag$ & \phantom{-}\textbf{0.397}{\scriptsize$^{***}$} & \phantom{-}0.106 & . & . & \phantom{-}0.196 & . & \phantom{-}0.149 & \phantom{-}0.104 & . & . \\
    Gemini 3 Flash$^\triangle$ & \phantom{-}\textbf{0.194}{\scriptsize$^{*}$} & \phantom{-}0.186 & . & . & . & . & . & \phantom{-}0.104 & . & . \\
    Gemini 3.5 Flash$^\dag$ & \phantom{-}\textbf{0.514}{\scriptsize$^{***}$} & \phantom{-}\textbf{0.290}{\scriptsize$^{**}$} & . & . & \phantom{-}0.137 & \phantom{-}0.126 & \phantom{-}\textbf{0.309}{\scriptsize$^{*}$} & \phantom{-}\textbf{0.211}{\scriptsize$^{*}$} & \phantom{-}0.177 & \phantom{-}0.135 \\
    Gemini 3.5 Flash$^\triangle$ & \phantom{-}\textbf{0.210}{\scriptsize$^{*}$} & \phantom{-}0.106 & \phantom{-}0.113 & . & \phantom{-}0.137 & \phantom{-}0.089 & \phantom{-}0.149 & \phantom{-}0.147 & \phantom{-}0.218 & \phantom{-}0.135 \\
    Qwen3 Omni$^\dag$ & . & . & . & . & . & . & . & . & . & . \\
    Qwen3 Omni$^\triangle$ & . & . & . & . & . & . & . & -0.104 & . & . \\
    Step Audio 2 Mini$^\dag$ & . & . & . & . & . & . & . & . & . & . \\
    Step Audio 2 Mini$^\triangle$ & . & . & \phantom{-}0.095 & . & . & . & \phantom{-}0.149 & . & . & . \\
    \midrule
    \multicolumn{11}{c}{C3 — Single-dimension focus; one dimension at a time} \\
    \midrule
    Gemini 3 Flash$^\dag$ & \phantom{-}\textbf{0.366}{\scriptsize$^{***}$} & \phantom{-}0.000 & \phantom{-}0.083 & . & \phantom{-}\textbf{0.319}{\scriptsize$^{*}$} & \phantom{-}0.089 & \phantom{-}0.213 & \phantom{-}0.061 & . & . \\
    Gemini 3 Flash$^\triangle$ & \phantom{-}\textbf{0.261}{\scriptsize$^{**}$} & \phantom{-}\textbf{0.267}{\scriptsize$^{*}$} & \phantom{-}0.000 & . & \phantom{-}0.055 & \phantom{-}0.021 & \phantom{-}0.213 & -0.054 & . & \phantom{-}0.135 \\
    Gemini 3.5 Flash$^\dag$ & \phantom{-}\textbf{0.355}{\scriptsize$^{***}$} & \phantom{-}0.178 & . & . & \phantom{-}0.243 & \phantom{-}\textbf{0.181}{\scriptsize$^{*}$} & \phantom{-}\textbf{0.459}{\scriptsize$^{**}$} & -0.060 & \phantom{-}\textbf{0.286}{\scriptsize$^{*}$} & \phantom{-}0.135 \\
    Gemini 3.5 Flash$^\triangle$ & \phantom{-}\textbf{0.194}{\scriptsize$^{*}$} & \phantom{-}0.170 & -0.155 & \phantom{-}0.100 & \phantom{-}\textbf{0.276}{\scriptsize$^{*}$} & \phantom{-}\textbf{0.232}{\scriptsize$^{**}$} & \phantom{-}\textbf{0.303}{\scriptsize$^{*}$} & \phantom{-}0.043 & \phantom{-}0.218 & -0.135 \\
    Qwen3 Omni$^\dag$ & . & . & . & . & \phantom{-}0.137 & \textbf{-0.184}{\scriptsize$^{*}$} & . & \phantom{-}0.026 & . & \phantom{-}0.165 \\
    Qwen3 Omni$^\triangle$ & \phantom{-}0.136 & \phantom{-}\textbf{0.243}{\scriptsize$^{*}$} & . & . & \phantom{-}\textbf{0.298}{\scriptsize$^{*}$} & \phantom{-}0.050 & -0.149 & \textbf{-0.214}{\scriptsize$^{*}$} & . & \phantom{-}0.139 \\
    Step Audio 2 Mini$^\dag$ & . & . & . & . & . & . & . & \phantom{-}0.104 & . & \phantom{-}0.000 \\
    Step Audio 2 Mini$^\triangle$ & . & -0.108 & . & \phantom{-}0.100 & . & . & . & \phantom{-}0.069 & . & \phantom{-}0.038 \\
    \bottomrule
  \end{tabular}%
  }

  \vspace{4pt}
  \parbox{\textwidth}{\footnotesize
    \textbf{Columns:} Phon.\,=\,Phonetic Accuracy; Lex.\,=\,Lexical Stress; Inton.\,=\,Intonation; Pros.S\,=\,Prosodic Stress; Pros.B\,=\,Prosodic Boundary; Rate\,=\,Speech Rate; Emo.\,=\,Emotional Appropriateness; Expr.\,=\,Expressiveness; Spk.\,=\,Speaker Identity; Hum.\,=\,Human Plausibility.
    \textbf{Transcript:}
    $^\dag$\,=\,with transcript; $^\triangle$\,=\,without transcript.
    \textbf{Bold} = significant ($p < .05$). $^{*}p<.05$, $^{**}p<.01$, $^{***}p<.001$ (two-sided).
    \textbf{Inter-annotator agreement:} $^\S$ = Krippendorff's $\alpha$<0.667 (see Table 3).}
\end{table*}

\paragraph{MOS Predictors and AudioLLM Judges show sensitivity to different dimensions.}
MOS predictors and \audiollmsingular~judges under a basic naturalness prompt (C1) attend to fundamentally different aspects of speech quality. Neither class aligns reliably with human judgments across the full set of perceptual dimensions. We observe that quality degradations in many dimensions go unnoticed by these models. These results constitute a controlled upper-bound diagnostic using injected synthetic errors; generalization to deployed TTS output remains uncertain. However, if naturally occurring errors are more subtle, this may represent a conservative estimate.
 
Every MOS predictor reaches significance on at least one dimension, with their attention concentrating on prosodic and paralinguistic dimensions generated via acoustic manipulation, while remaining insensitive to word-level dimensions (see Table~\ref{tab:tau_bridge}). Human Plausibility, constructed via inserted glitches, is the most consistently detected dimension, reaching significance across all tested models. Speech Rate Appropriateness, constructed via duration scaling, is also significant across 6 of the 8 models. Both patterns are consistent with MOS predictors tracking signal-level degradation, reflecting their origins in telephony-era voice quality assessment where such artifacts were the primary target. Expressiveness also reaches significance for six of the eight models, though whether this reflects genuine sensitivity to expressive delivery or a response to shared low-level acoustic properties across the Expressiveness, Speech Rate, and Human Plausibility stimuli remains unclear.  In contrast, no MOS models show significant detection of Phonetic Accuracy, Lexical Stress, and Prosodic Boundary Placement, dimensions where TTS systems produce linguistically meaningful errors that human listeners reliably detect, but MOS predictors do not.

To confirm that observed patterns are not driven by benchmark noise, we perform the same analysis on a balanced-unanimous subset of 516 samples where all three annotators agreed (Appendix Table~\ref{tab:unanimous-robustness}) . The main patterns remain unchanged. MOS predictors are still sensitive to prosodic- and paralinguistic- but not word-level dimensions; in contrast, AudioLLM judges have scattered sensitivity, and Gemini remains sensitive to word-level errors when the transcript is provided. On lower-agreement dimensions, some correlations become stronger. For Expressiveness (\(\alpha=0.46\)), AudioBox\(_{\mathrm{CU}}\) rises from \(0.42\) to \(0.67\), and NISQA from \(0.35\) to \(0.64\). This suggests that annotation noise was weakening real alignment rather than creating spurious correlations. Because unanimous items may contain more obvious errors, we treat this as a robustness check rather than a revised estimate of model performance.

\paragraph{Some significant correlations are negative.} We observe a misalignment between some predictors and human perceptual judgments. AudioBox$_\mathrm{PC}$ shows a strong negative relationship with Human Plausibility; negative effects also appear across AudioBox$_\mathrm{CE}$,  AudioBox$_\mathrm{PQ}$ and DNSMOS Pro$_\mathrm{BVCC}$ on Speaker Identity Consistency, Prosodic Stress and Speech Rate. We hypothesize that these predictors assign higher scores to the controlled acoustic distortions applied to the paralinguistic dimensions (see Appendix~\ref{app:praat}), rating degraded samples more favorably than human listeners do. 

\paragraph{Generic naturalness prompts yield sparse, inconsistent sensitivity.} In contrast to MOS models, \audiollmsingular~judges prompted for a generic naturalness rating (C1) show sparse and scattered attention to human-perceptible dimensions (see Table~\ref{tab:tau_bridge}). Gemini 3.5 Flash with transcript shows the broadest coverage of any \audiollm~condition, reaching significance on dimensions spanning the word-level, prosodic, and paralinguistic tiers, while every other model and condition reaches at most two or three significant dimensions. Beyond the Gemini family, sensitivity under C1 is sparse to absent, with no dimension reliably detected across more than one model, and the few significant effects are directionally inconsistent. Taken together, the C1 baseline shows selective sensitivity across the Gemini models, while the other models show little evidence of reliable sensitivity to any dimension under an underspecified naturalness prompt.

\paragraph{Schema guidance helps selectively.} 

We find that schema guidance recovers word-level sensitivity absent under the unguided naturalness prompt (C1): providing the full \numattributes-dimension schema while maintaining a single aggregated score (C2a) increases sensitivity to word-level dimensions for models that showed none under C1, suggesting that this capacity was present but not elicited without schema guidance (see Table~\ref{tab:tau_lalm_c2c3}).

The clearest gain appears in the no-transcript condition: both Gemini variants show no significant word-level correlations under the basic naturalness prompt (C1) but achieve statistically significant correlations when provided the detailed schema (C2a). This supports schema guidance improving model sensitivity on word-level dimensions, and underscores how under-specified a bare naturalness prompt is for surfacing sensitivity to which the model evidently has access.

However, this benefit does not extend uniformly across all dimensions: some correlations significant under C1 lose significance under C2a, while new ones emerge. Paralinguistic dimensions remain largely resistant across all conditions, with significant correlations sparse and inconsistent.

Output collapse (where a model predicts the same score for every sample, leaving correlation undefined) is also observed under C2a; for example, Step Audio 2 Mini assigns the maximum score to every sample regardless of true label.

When models are provided the schema and also required to score each dimension independently (C2b), output collapse becomes the dominant failure mode. Two of the four models assign the maximum score across nearly all dimensions and both transcript conditions, leaving correlation undefined in almost all cells.

\paragraph{Transcript access does not consistently increase sensitivity to any dimension.}
Transcript availability produces inconsistent effects across conditions and models, with no clean directional pattern. The models that show sensitivity to word-level dimensions under the unguided naturalness prompt C1 (Gemini 3 Flash and Gemini 3.5 Flash) only do so when the transcript is provided. However, once the schema is provided (C2a), this dependency does not persist; both Gemini variants become more significant on phonetic accuracy with the transcript.

When models are prompted for each dimension independently (C3), both Gemini variants remain significant on Phonetic Accuracy with stronger effect sizes when the transcript is provided. For paralinguistic dimensions, Gemini 3.5 Flash also reaches significance on Emotional Appropriateness in both transcript conditions. However, the pattern runs the opposite way for paralinguistic dimensions for two other models; Qwen3 Omni and Step Audio 2 Mini show significant Speaker Identity sensitivity under C2a only with the transcript, and show no sensitivity to the other paralinguistic dimensions in either condition.

The absence of a consistent transcript benefit suggests that current \audiollm~judges are not consistent in their ability to leverage textual grounding to improve dimension-specific detection, even for dimensions such as Phonetic Accuracy where the reference transcript is directly informative.

\paragraph{\audiollmsingular~sensitivity remains inconsistent under per-dimension prompting.} Under C3 (per-dimension prompting), sensitivity remains sparse and inconsistent, with a pattern that largely differs from C1 and C2a. An exception is Gemini's increased sensitivity to Phonetic Accuracy under transcript provision, which holds across prompt variations. Negative correlations still exist for some models on isolated dimensions. Step Audio 2 Mini shows no significant correlation under any transcript condition in C3, consistent with its performance across all other conditions and suggesting a fundamental sensitivity limitation rather than a prompting artifact.
\section{Conclusion}
We present the first dimension-level meta-evaluation benchmark for TTS, decomposing naturalness into \numattributes~linguistically grounded perceptual attributes annotated by trained linguists. Across four MOS predictors and four \audiollm~judges, we find that current automated evaluators exhibit systematic but distinct blind spots: MOS predictors align strongly with signal-level artifacts but fail on word-level dimensions, whereas \audiollm~judges show selective, prompt-dependent detection that never generalizes across the full dimension set.

We further show that the measured performance of \audiollmsingular~judges is substantially affected by the way that they are prompted: we observed that providing schema guidance improves sensitivity for some dimensions and models, while asking models to score each dimension individually often degrades output through prediction collapse.

Taken together, our findings suggest that naturalness cannot be treated as a single scalar construct for modern TTS evaluation and that robust evaluation will require dimension-aware benchmarks.
\section*{Limitations}

\textbf{Ecological validity.} The dataset is not intended to represent a corpus of spontaneous TTS failures observed in real-world use, but rather consists of controlled synthetic-speech stimuli containing perceptually salient errors introduced through targeted text prompting and acoustic manipulation in Praat. This design enables systematic coverage of specific perceptual dimensions while prioritizing experimental control over resemblance to naturally occurring TTS behavior; accordingly, the results should be interpreted as an upper-bound estimate of detection sensitivity under known error conditions, rather than as representative evidence about TTS failures in the wild.

\noindent \textbf{Stimulus complexity ceiling.} Harvard Sentences average 7--8 words, contain few idioms or proper nouns, and consist of isolated sentences with no discourse context. This makes them ideal for eliciting clear word-level failures but limits the generalizability of the findings to more complex speech involving multiple speakers, multiple turns, or longer contexts, in which the acceptability of prosodic and paralinguistic features may be influenced by discourse context.

\noindent \textbf{Single architecture for IPA-controlled generation.} The formatting and reliability of phoneme-level input vary widely across providers. Our pipeline currently supports only Cartesia Sonic; therefore, all IPA-driven samples reflect a single TTS architecture. We acknowledge that evaluation across multiple systems remains an important direction, and our publicly released generation pipeline is designed to facilitate this extension.

\noindent \textbf{Annotation scale granularity.} Binary scales do not capture full fine-grained quality gradients. In contrast, ordinal scales provide granularity but also the potential for response biases and inconsistent boundary interpretation between levels. In this context, binary scales were a deliberate scope choice. Rather than full perceptual grading, our benchmark targets error detection.

\noindent \textbf{Annotation reliability.} Majority-class downsampling to achieve class balance reduces effective sample sizes, most consequentially for Emotional Appropriateness (n=46) and Lexical Stress (n=90), where agreement estimates carry wider confidence intervals.

\section*{Ethics Statement}
Our proposed benchmark is intended solely to support more targeted and interpretable TTS evaluation research and is not intended for discriminatory applications or commercial ranking of TTS providers. All annotators are first-language English speakers and trained linguists who participated voluntarily at standard rate compensation. No sensitive or personally identifiable data was collected.

All of our speech samples are synthetically generated or acoustically manipulated; no real human speech recordings were collected. Source corpora and APIs are used in accordance with their respective licenses and terms of service (see Appendix~\ref{app:sec_license}).

Regarding Large Language Model (LLM) usage in manuscript preparation, we utilize them  solely to refine the language used in paper to improve clarity and correctness, without generating any substantial content or claims.
\bibliography{custom}

@inproceedings{geneva23_interspeech,
  title     = {{Accentor: An Explicit Lexical Stress Model for TTS Systems}},
  author    = {Diana Geneva and Georgi Shopov and Kostadin Garov and Maria Todorova and Stefan Gerdjikov and Stoyan Mihov},
  year      = {2023},
  booktitle = {{Interspeech 2023}},
  pages     = {4848--4852},
  doi       = {10.21437/Interspeech.2023-433},
  issn      = {2958-1796},
}

@inproceedings{xie2025towards,
  title={Towards controllable speech synthesis in the era of large language models: A systematic survey},
  author={Xie, Tianxin and Rong, Yan and Zhang, Pengfei and Wang, Wenwu and Liu, Li},
  booktitle={Proceedings of the 2025 Conference on Empirical Methods in Natural Language Processing},
  pages={764--791},
  year={2025}
}

@article{tan2021survey,
  title={A survey on neural speech synthesis},
  author={Tan, Xu and Qin, Tao and Soong, Frank and Liu, Tie-Yan},
  journal={arXiv preprint arXiv:2106.15561},
  year={2021}
}

@article{ulgen2026objective,
  title={Objective Evaluation of Prosody and Intelligibility in Speech Synthesis via Conditional Prediction of Discrete Tokens},
  author={Ulgen, Ismail Rasim and Du, Zongyang and Lu, Junchen and Koehn, Philipp and Sisman, Berrak},
  journal={IEEE Open Journal of Signal Processing},
  year={2026},
  publisher={IEEE}
}

@inproceedings{baughan2023mixed,
  title={A mixed-methods approach to understanding user trust after voice assistant failures},
  author={Baughan, Amanda and Wang, Xuezhi and Liu, Ariel and Mercurio, Allison and Chen, Jilin and Ma, Xiao},
  booktitle={Proceedings of the 2023 CHI Conference on Human Factors in Computing Systems},
  pages={1--16},
  year={2023}
}

@inproceedings{gutierrez2021location,
  title={Location, Location: Enhancing the Evaluation of Text-to-Speech synthesis using the Rapid Prosody Transcription Paradigm},
  author={Gutierrez, Elijah and Oplustil-Gallegos, Pilar and Lai, Catherine},
  booktitle={Proc. SSW 2021},
  pages={25--30},
  year={2021}
}

@inproceedings{patton2016automos,
title	= {AutoMOS: Learning a non-intrusive assessor of naturalness-of-speech},author	= {Brian Patton and Yannis Agiomyrgiannakis and Michael Terry and Kevin Wilson and Rif A. Saurous and D. Sculley},year	= {2016},URL	= {https://arxiv.org/abs/1611.09207},booktitle	= {NIPS 2016 End-to-end Learning for Speech and Audio Processing Workshop}}

@inproceedings{huang2024voicemos,
  title={The VoiceMOS challenge 2024: Beyond speech quality prediction},
  author={Huang, Wen-Chin and Fu, Szu-Wei and Cooper, Erica and Zezario, Ryandhimas E and Toda, Tomoki and Wang, Hsin-Min and Yamagishi, Junichi and Tsao, Yu},
  booktitle={2024 IEEE Spoken Language Technology Workshop (SLT)},
  pages={803--810},
  year={2024},
  organization={IEEE}
}

@inproceedings{huang2022voicemos,
  title={The VoiceMOS Challenge 2022},
  author={Huang, Wen Chin and Cooper, Erica and Tsao, Yu and Wang, Hsin-Min and Toda, Tomoki and Yamagishi, Junichi},
  booktitle={Proc. Interspeech 2022},
  pages={4536--4540},
  year={2022}
}

@article{mittag2021nisqa,
  title={NISQA: A deep CNN-self-attention model for multidimensional speech quality prediction with crowdsourced datasets},
  author={Mittag, Gabriel and Naderi, Babak and Chehadi, Assmaa and M{\"o}ller, Sebastian},
  journal={arXiv preprint arXiv:2104.09494},
  year={2021}
}

@inproceedings{saeki2022utmos,
  title={UTMOS: UTokyo-SaruLab System for VoiceMOS Challenge 2022},
  author={Saeki, Takaaki and Xin, Detai and Nakata, Wataru and Koriyama, Tomoki and Takamichi, Shinnosuke and Saruwatari, Hiroshi},
  booktitle={Proc. Interspeech 2022},
  pages={4521--4525},
  year={2022}
}

@article{tjandra2025meta,
    title={Meta audiobox aesthetics: Unified automatic quality assessment for speech, music, and sound},
  author={Tjandra, Andros and Wu, Yi-Chiao and Guo, Baishan and Hoffman, John and Ellis, Brian and Vyas, Apoorv and Shi, Bowen and Chen, Sanyuan and Le, Matt and Zacharov, Nick and others},
  journal={arXiv preprint arXiv:2502.05139},
  year={2025}
}

@inproceedings{baba2024utmosv2,
  title     = {The T05 System for The {V}oice{MOS} {C}hallenge 2024: Transfer Learning from Deep Image Classifier to Naturalness {MOS} Prediction of High-Quality Synthetic Speech},
  author    = {Baba, Kaito and Nakata, Wataru and Saito, Yuki and Saruwatari, Hiroshi},
  booktitle = {IEEE Spoken Language Technology Workshop (SLT)},
  year      = {2024},
}

@inproceedings{cumlin24_interspeech,
  title     = {{DNSMOS Pro: A Reduced-Size DNN for Probabilistic MOS of Speech}},
  author    = {Fredrik Cumlin and Xinyu Liang and Victor Ungureanu and Chandan {K. A. Reddy} and Christian Schüldt and Saikat Chatterjee},
  year      = {2024},
  booktitle = {{Interspeech 2024}},
  pages     = {4818--4822},
  doi       = {10.21437/Interspeech.2024-478},
  issn      = {2958-1796},
}

@inproceedings{reddy2021dnsmos,
  title={DNSMOS: A non-intrusive perceptual objective speech quality metric to evaluate noise suppressors},
  author={Reddy, Chandan KA and Gopal, Vishak and Cutler, Ross},
  booktitle={ICASSP 2021-2021 IEEE International Conference on Acoustics, Speech and Signal Processing (ICASSP)},
  pages={6493--6497},
  year={2021},
  organization={IEEE}
}

@inproceedings{minixhofer2026ttsds2,
  title={TTSDS2: resources and benchmark for evaluating human-quality text to speech systems},
  author={Minixhofer, Christoph and Klejch, Ondrej and Bell, Peter},
  booktitle={International Conference on Learning Representations},
  volume={2026},
  pages={29228--29247},
  year={2026}
}

@article{zhang2025speechjudge,
  title={SpeechJudge: Towards Human-Level Judgment for Speech Naturalness},
  author={Zhang, Xueyao and Wang, Chaoren and Liao, Huan and Li, Ziniu and Wang, Yuancheng and Wang, Li and Jia, Dongya and Chen, Yuanzhe and Li, Xiulin and Chen, Zhuo and others},
  journal={arXiv preprint arXiv:2511.07931},
  year={2025}
}

@inproceedings{manakul2026audiojudge,
  title={Audiojudge: Understanding what works in large audio model based speech evaluation},
  author={Manakul, Potsawee and Gan, Woody Haosheng and Ryan, Michael J and Khan, Ali Sartaz and Sirichotedumrong, Warit and Pipatanakul, Kunat and Held, William Barr and Yang, Diyi},
  booktitle={Proceedings of the 19th Conference of the European Chapter of the Association for Computational Linguistics (Volume 1: Long Papers)},
  pages={3644--3663},
  year={2026}
}

@inproceedings{chiang-etal-2025-audio,
    title = "Audio-Aware Large Language Models as Judges for Speaking Styles",
    author = "Chiang, Cheng-Han  and
      Wang, Xiaofei  and
      Lin, Chung-Ching  and
      Lin, Kevin  and
      Li, Linjie  and
      Kopetz, Radu  and
      Qian, Yao  and
      Wang, Zhendong  and
      Yang, Zhengyuan  and
      Lee, Hung-yi  and
      Wang, Lijuan",
    editor = "Christodoulopoulos, Christos  and
      Chakraborty, Tanmoy  and
      Rose, Carolyn  and
      Peng, Violet",
    booktitle = "Findings of the Association for Computational Linguistics: EMNLP 2025",
    month = nov,
    year = "2025",
    address = "Suzhou, China",
    publisher = "Association for Computational Linguistics",
    url = "https://aclanthology.org/2025.findings-emnlp.25/",
    doi = "10.18653/v1/2025.findings-emnlp.25",
    pages = "467--480",
    ISBN = "979-8-89176-335-7",
}

@article{monjur2025speechqualityllm,
  title={SpeechQualityLLM: LLM-Based Multimodal Assessment of Speech Quality},
  author={Monjur, Mahathir and Nirjon, Shahriar},
  journal={arXiv preprint arXiv:2512.08238},
  year={2025}
}

@article{lee2026speakersleuth,
  title={SpeakerSleuth: Evaluating Large Audio-Language Models as Judges for Multi-turn Speaker Consistency},
  author={Lee, Jonggeun and Pyo, Junseong and Seo, Gyuhyeon and Jo, Yohan},
  journal={arXiv preprint arXiv:2601.04029},
  year={2026}
}

@article{manku2026emergenttts,
  title={Emergenttts-eval: Evaluating tts models on complex prosodic, expressiveness, and linguistic challenges using model-as-a-judge},
  author={Manku, Ruskin Raj and Tang, Yuzhi and Shi, Xingjian and Li, Mu and Smola, Alexander},
  journal={Advances in Neural Information Processing Systems},
  volume={38},
  year={2026}
}

@article{huang2025instructttseval,
  title={Instructttseval: Benchmarking complex natural-language instruction following in text-to-speech systems},
  author={Huang, Kexin and Tu, Qian and Fan, Liwei and Yang, Chenchen and Zhang, Dong and Li, Shimin and Fei, Zhaoye and Cheng, Qinyuan and Qiu, Xipeng},
  journal={arXiv preprint arXiv:2506.16381},
  year={2025}
}

@inproceedings{kirkland23_ssw,
  title     = {{Stuck in the MOS pit: A critical analysis of MOS test methodology in TTS evaluation}},
  author    = {Ambika Kirkland and Shivam Mehta and Harm Lameris and Gustav Eje Henter and Eva Szekely and Joakim Gustafson},
  year      = {2023},
  booktitle = {{12th ISCA Speech Synthesis Workshop (SSW2023)}},
  pages     = {41--47},
  doi       = {10.21437/SSW.2023-7},
}

@inproceedings{vallesperez21_interspeech,
  title     = {{Improving Multi-Speaker TTS Prosody Variance with a Residual Encoder and Normalizing Flows}},
  author    = {Iván Vallés-Pérez and Julian Roth and Grzegorz Beringer and Roberto Barra-Chicote and Jasha Droppo},
  year      = {2021},
  booktitle = {{Interspeech 2021}},
  pages     = {3131--3135},
  doi       = {10.21437/Interspeech.2021-562},
  issn      = {2958-1796},
}

@book{crystal1969prosodic,
  author    = {David Crystal},
  title     = {Prosodic Systems and Intonation in {English}},
  year      = {1969},
  publisher = {Cambridge University Press},
  address   = {Cambridge}
}

@article{jesse2017english,
  title={English listeners use suprasegmental cues to lexical
         stress early during spoken-word recognition},
  author={Jesse, Alexandra and Poellmann, Katja and Kong, Ying-Yee},
  journal={Journal of Speech, Language, and Hearing Research},
  volume={60},
  number={1},
  pages={190--198},
  year={2017},
  doi={10.1044/2016_JSLHR-H-15-0340}
}

@article{10.1613/jair.1.12752,
author = {Uma, Alexandra N. and Fornaciari, Tommaso and Hovy, Dirk and Paun, Silviu and Plank, Barbara and Poesio, Massimo},
title = {Learning from Disagreement: A Survey},
year = {2022},
issue_date = {Jan 2022},
publisher = {AI Access Foundation},
address = {El Segundo, CA, USA},
volume = {72},
issn = {1076-9757},
url = {https://doi.org/10.1613/jair.1.12752},
doi = {10.1613/jair.1.12752},
journal = {J. Artif. Int. Res.},
month = jan,
pages = {1385–1470},
numpages = {86}
}

@article{ieee1969harvard,
  title={{IEEE} recommended practice for speech quality measurements},
  author={{Institute of Electrical and Electronics Engineers}},
  journal={IEEE Transactions on Audio and Electroacoustics},
  volume={17},
  number={3},
  pages={225--246},
  year={1969},
  note={IEEE Std 297-1969. Appendix C: 1965 Revised List of Phonetically Balanced Sentences (Harvard Sentences)}
}

@techreport{krippendorff2011computing,
  author      = {Krippendorff, Klaus},
  title       = {Computing {Krippendorff}'s Alpha-Reliability},
  year        = {2011},
  institution = {University of Pennsylvania, Annenberg School for Communication},
  address     = {Philadelphia, PA},
  type        = {Departmental Papers},
  url         = {https://repository.upenn.edu/entities/publication/034a6030-c584-4d14-9d3d-7b7e8d16df20}
}

@article{boersma2001praat,
  author   = {Boersma, Paul},
  title    = {Praat, a system for doing phonetics by computer},
  journal  = {Glot International},
  year     = {2001},
  volume   = {5},
  number   = {9/10},
  pages    = {341--345},
}

@article{jadoul2018introducing,
  author    = {Jadoul, Yannick and Thompson, Bill and de Boer, Bart},
  title     = {Introducing {Parselmouth}: A {Python} Interface to {Praat}},
  journal   = {Journal of Phonetics},
  volume    = {71},
  pages     = {1--15},
  year      = {2018},
  doi       = {10.1016/j.wocn.2018.07.001}
}

@misc{cartesia2025sonic3,
  author       = {{Cartesia AI}},
  title        = {Sonic-3: Streaming Text-to-Speech Model},
  year         = {2025},
  month        = {October},
  howpublished = {\url{https://docs.cartesia.ai/build-with-cartesia/tts-models/latest}},
  note         = {Accessed: 2025}
}
\UseRawInputEncoding

\section{Appendix}
\label{sec:appendix}

\section*{Glossary of Terms}
\addcontentsline{toc}{section}{Glossary of Terms}


\subsubsection*{Linguistic Theory}

\begin{description}[
  leftmargin=0pt,
  labelwidth=0pt,
  labelsep=0pt,
  itemindent=0pt,
  itemsep=3pt,
  parsep=0pt,
  topsep=2pt,
  font=\normalfont\bfseries
]

  \item[Acoustic.]~Of or relating to the physical properties of sound as a
    mechanical wave. In speech science, acoustic analysis involves
    measurements of frequency, amplitude, duration, and spectral
    characteristics derived from the audio signal.

  \item[Lexical.]~Of or relating to the words or vocabulary of a language,
    as distinct from its grammatical structure. In speech, lexical properties
    include word identity, lexical stress assignment, and phonological form.

  \item[Linguistic.]~Of or relating to language or linguistics. In speech
    evaluation, pertaining to features conventionally encoded in the grammar
    of a language, including segmental and prosodic structure.

  \item[Phoneme.]~The smallest contrastive unit of sound capable of
    distinguishing meaning. Phonemes are abstract categories realized by
    phonetically similar variants (allophones) whose distribution is
    governed by phonological context.

  \item[Phonetics.]~The branch of linguistics concerned with the physical
    and perceptual properties of speech sounds.

  \item[Phonology.]~The branch of linguistics concerned with the abstract,
    rule-governed organization of sound systems. Phonology describes the
    systematic patterns governing the distribution and combination of
    phonemes.

  \item[Segmental.]~Pertaining to individual consonant and vowel segments
    comprising the phonemic inventory of a language. Segmental quality
    encompasses place and manner of articulation, voicing, vowel quality,
    and allophonic processes such as aspiration and flapping.

  \item[Semantic.]~Of or relating to meaning in language. In TTS
    evaluation, semantic considerations include whether prosodic
    realization---particularly intonation and focus---accurately reflects
    information structure, such as given/new status and contrastive
    emphasis.

  \item[Suprasegmental.]~Pertaining to phonological features spanning
    units larger than a single segment: stress, pitch accent, intonation,
    rhythm, and boundary placement. Also termed \textit{prosodic} features.

  \item[Syntactic.]~Of or relating to the grammatical structure and word order of
    sentences. Syntactic constituent boundaries frequently align with
    prosodic boundaries, and syntactic relationships influence prominence
    placement and intonational contour shape.

    \item[Paralinguistic.]~Pertaining to communicative cues that accompany linguistic expression without constituting part of its lexical, grammatical, or phonological structure. Such cues may convey affect, stance, or speaker identity. Although paralinguistic features may be vocal, visual, gestural, or otherwise embodied, this glossary uses the term in the narrower context of TTS evaluation to refer only to vocal cues.

\end{description}


\subsubsection*{Notational}

\begin{description}[
  leftmargin=0pt,
  labelwidth=0pt,
  labelsep=0pt,
  itemindent=0pt,
  itemsep=3pt,
  parsep=0pt,
  topsep=2pt,
  font=\normalfont\bfseries
]

  \item[f\textsubscript{0} (Fundamental Frequency)]~The lowest frequency
    component of a voiced speech waveform, corresponding to the rate of
    vocal fold vibration. f\textsubscript{0} is the primary acoustic
    correlate of perceived pitch and a key parameter of intonation and
    stress.

  \item[IPA (International Phonetic Alphabet)]~A standardized phonetic
    notation system maintained by the International Phonetic Association,
    providing a unique symbol for every sound in any human language. Used
    for both broad (phonemic) and narrow (phonetic) transcription.

\end{description}

\subsubsection*{Speech \& Language Technology}

\begin{description}[
  leftmargin=0pt,
  labelwidth=0pt,
  labelsep=0pt,
  itemindent=0pt,
  itemsep=3pt,
  parsep=0pt,
  topsep=2pt,
  font=\normalfont\bfseries
]

  \item[LALM (Large Audio Language Model)]~A multimodal extension of
    large language models incorporating audio as an input and/or output
    modality. LALMs can process and generate spoken language, in many
    architectures subsuming TTS and ASR within a unified model.

  \item[LLM (Large Language Model)]~A neural language model trained on
    large-scale text corpora via self-supervised objectives, capable of
    generation, classification, and reasoning across a wide range of
    text-based tasks.

  \item[TTS (Text-to-Speech)]~A class of speech synthesis technology
    converting written text to spoken audio. Modern neural TTS systems
    leverage sequence-to-sequence architectures and neural vocoders trained
    at scale to produce high-fidelity synthetic speech.

  \item[MOS (Mean Opinion Score)]~A perceptual evaluation metric in which
    listeners rate stimulus quality on a scale of 1--5; the MOS is the
    arithmetic mean of collected ratings. Originally standardized for
    telephony (ITU-T P.800), MOS and its variants are standard in TTS
    evaluation.

\end{description}


\subsection{Data Scale}
\subsubsection{Scale Collapse}
 For most ternary dimensions, collapsed
$\alpha$ is comparable to raw ordinal $\alpha$; for expressiveness, collapsed
$\alpha$ exceeds raw $\alpha$ , indicating that the middle
level adds noise rather than signal. Three-way splits on ternary dimensions
(each rater assigning a distinct level) occur only on the 1-versus-2 boundary
and are irresolvable by majority vote; these samples are excluded from the
final dataset for that dimension.

\begin{figure*}[t]
  \centering
  \includegraphics[trim={0.0cm 0.0cm 0.0cm 2.0cm}, clip, width=1.0\textwidth]{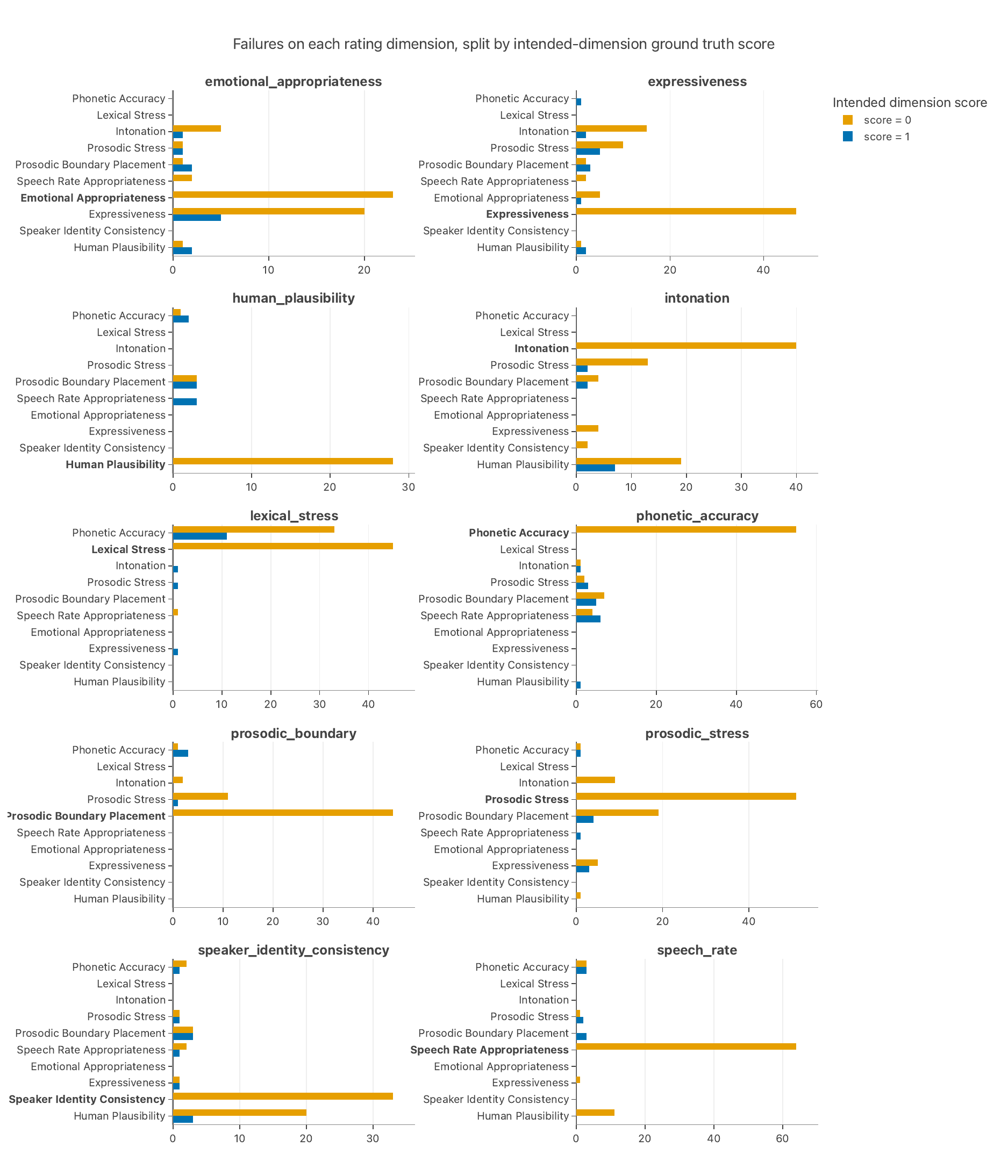}
  \caption{\textbf{Co-occurrence of perceptual failures across speech quality dimensions.} Per-dimension failure counts stratified by intended-dimension ground-truth label. Each subplot corresponds to one intended dimension, i.e., the dimension for which the sample was analyzed. Bars show the number of score-0 (failure) observations on each rating dimension (y-axis), split by whether the sample's ground-truth label on its \textit{own} intended dimension is a failure (\textcolor{orange}{orange}, score~=~0) or a pass (\textcolor{blue}{blue}, score~=~1). Tall orange bars on dimensions \textit{other} than the intended one indicate co-occurrence of errors.
  Tall blue bars on dimensions \textit{other} than the intended one indicate samples that pass on their intended construct but nonetheless exhibit failures on unrelated dimensions. For a TTS evaluator with perfect sensitivity across dimensions, these unrelated-dimension errors would appear as false positives.}
  \label{fig:correlation_by_dimension}
\end{figure*}


\onecolumn
\begin{longtable}{p{3.5cm} p{1.0cm} p{\dimexpr\textwidth-3.5cm-1.0cm-6\tabcolsep-2\arrayrulewidth\relax}}
\caption{\textbf{Annotation rubric for all ten speech quality dimensions.} 
Each dimension is described in full in \S\ref{subsec:evaluation_schema}. All ternary dimensions are collapsed to binary scoring as described in Appendix\ref{app:human_annotations}.}
\label{tab:full_rubric} \\
\toprule
\textbf{Dimension} & \textbf{Score} & \textbf{Meaning} \\
\midrule
\endfirsthead

\multicolumn{3}{l}{\small\textit{(Continued from previous page)}} \\
\toprule
\textbf{Dimension} & \textbf{Score} & \textbf{Meaning} \\
\midrule
\endhead

\midrule
\multicolumn{3}{r}{\small\textit{(Continued on next page)}} \\
\endfoot

\bottomrule
\endlastfoot

\multirow{3}{3.5cm}{\textbf{Phonetic Accuracy} (\S\ref{app:guide_phonetic_accuracy})}
  & 1 & One or more consonants or vowels are wrong and the error(s) are very
        noticeable. You might mishear the word entirely. \\[3pt]
  & 2 & One or more consonants or vowels are slightly off, but you can still
        tell what word was intended without much effort. \\[3pt]
  & 3 & All consonants and vowels sound correct. No perceivable phonetic
        errors. \\
\midrule

\multirow{2}{3.5cm}{\textbf{Lexical Stress} (\S\ref{app:guide_lexical_stress})}
  & 0 & At least one word has stress on the wrong syllable. Even one error
        earns a~0. \\[3pt]
  & 1 & Every multi-syllable word has stress on the correct syllable with
        appropriate vowel quality. No errors detected. \\
\midrule

\multirow{3}{3.5cm}{\textbf{Intonation} (\S\ref{app:guide_intonation})}
  & 1 & One or more pitch contour errors cause high negative impact. The
        intonation may confuse the intended sentence type or signal an
        unintended attitude. \\[3pt]
  & 2 & One or more errors cause low negative impact. Something about the
        melody sounds slightly unnatural but does not cause confusion. \\[3pt]
  & 3 & No perceivable errors. The intonation sounds natural and appropriate
        for the sentence and its context. \\
\midrule

\multirow{2}{3.5cm}{\textbf{Prosodic Stress} (\S\ref{app:guide_prosodic_stress})}
  & 0 & Prominence falls on words that do not warrant it, or words that
        should be prominent are not. \\[3pt]
  & 1 & The choice of prominent words sounds predictable and appropriate.
        New or important information is highlighted; given material is
        de-emphasized. \\
\midrule

\multirow{3}{3.5cm}{\textbf{Prosodic Boundary Placement} (\S\ref{app:guide_prosodic_boundary})}
  & 1 & A boundary is missing where it is needed or present where it should
        not be, with severe impact. You cannot recover the intended sentence
        structure while listening. \\[3pt]
  & 2 & A boundary is in the wrong position and noticeably hurts the
        experience, but you can still follow the sentence structure. \\[3pt]
  & 3 & Words are grouped into phrases that align naturally with the
        sentence's syntactic and semantic structure. \\
\midrule

\multirow{3}{3.5cm}{\textbf{Speech Rate Appropriateness} (\S\ref{app:guide_speech_rate})}
  & 1 & Rate is clearly inappropriate: so fast words blur together, or so
        slow the pace feels labored. \\[3pt]
  & 2 & Rate is slightly too fast, too slow, or too uniform, but you can
        follow the content without real difficulty. \\[3pt]
  & 3 & Tempo feels natural and varies appropriately. Complex material gets
        more time; parentheticals move briskly. \\
\midrule

\pagebreak[3]
\multirow{2}{3.5cm}{\textbf{Emotional Appropriateness} (\S\ref{app:guide_emotional_approrpriateness})}
  & 0 & The emotional tone is wrong for the content. The mismatch is clearly
        perceivable. \\[3pt]
  & 1 & The emotional tone is a plausible match for the text and context. \\
\midrule

\multirow{3}{3.5cm}{\textbf{Expressiveness} (\S\ref{app:guide_expressiveness})}
  & 1 & Delivery is extremely flat and robotic, or excessively exaggerated,
        to the point of seriously detracting from the experience. \\[3pt]
  & 2 & Delivery is slightly too flat or slightly too dramatic for the
        content, but not very distracting. \\[3pt]
  & 3 & Delivery has natural, appropriate variation; sounds engaged and
        dynamic without drawing attention to itself. \\
\midrule

\multirow{2}{3.5cm}{\textbf{Speaker Identity Consistency} (\S\ref{app:guide_speaker_identity})}
  & 0 & The voice changes in a way that makes it sound like a different
        person, either within or across sentences. \\[3pt]
  & 1 & The voice is recognizably consistent throughout. It sounds like a
        single individual, even with natural variation in expressiveness or
        loudness. \\
\midrule

\multirow{2}{3.5cm}{\textbf{Human Plausibility} (\S\ref{app:guide_human_plausibility})}
  & 0 & The audio contains at least one moment that could not have been
        produced by a human speaker: robotic or metallic artifacts,
        impossible pitch jumps, audio glitches, or unnaturally cut-off
        syllables. \\[3pt]
  & 1 & The entire audio stream sounds like it could have been produced by
        a human speaker. \\

\end{longtable}
\twocolumn

\subsection{Human Annotations}
\label{app:human_annotations}
\paragraph{Annotator Background} All annotators are first-language English speakers and trained linguists. Annotators participated voluntarily at the standard rate compensation. The annotation task involved listening to synthesized speech samples and marking perceptual quality dimensions based on the linguistic-verified guidelines; no sensitive or personal data was collected.

\paragraph{Annotator Iteration}
For dimensions that contain inherent human subjectivity, we initially tested capturing a severity gradient with a ternary scale (1-3). However, annotators could not reliably distinguish between non-perfect rating levels, so we collapsed these to binary, trading nuance for dataset reliability.

Since not all generations reliably exhibit the intended failure, we generated an excess of samples and relied on human annotation to establish ground-truth labels (see Table~\ref{tab:confusion}).

Annotation is conducted through a purpose-built interface presenting the audio, the
dimension definition, the revealable transcript, and the response options in a single
screen, minimizing cognitive load and reducing interface-induced variance between raters.

\paragraph{Annotation Guide}
The annotation guide provides detailed explanations and rubric ratings on how to rate synthetic (TTS) speech on each metric in
the evaluation schema. For every metric, you will find: a plain-language
explanation of what the metric captures, what to listen for, a rating scale
with clear descriptions of each score, and examples to anchor annotators' judgments.

The metrics are grouped into three dimensions: \textbf{Word-level} (are the
right sounds and stress patterns produced for each word?), \textbf{Prosodic}
(is the speech organized into phrases with appropriate melody, emphasis, and
timing?), and \textbf{Paralinguistic} (does the speech convey appropriate
emotion, expressiveness, and speaker characteristics?). The details are provided in Table \ref{tab:full_rubric}.

\paragraph{General Labeling Guideline}

Listen to each utterance at least twice before scoring. On the first pass, get
a general impression. On the second pass, focus on whichever metric you are
currently rating. Try not to let a strong error in one metric bias your rating
in another: a word might be mispronounced (a Phonetic Accuracy issue) while
the prosody is perfectly fine, or vice versa. Rate each metric independently.

\paragraph{Word-Level Metrics}

Word-level metrics ask: does each individual word sound right? This covers
whether the TTS system picked the correct sounds (phonemes) and placed stress
on the correct syllable within each word.

\subsubsection{Phonetic Accuracy}
\label{app:guide_phonetic_accuracy}
\textit{What you are listening for.} Does every word contain the right speech
sounds? This includes consonants, vowels, and coarticulation. You are checking
whether the system produced the correct phonemes for each word, especially for
tricky cases like heteronyms, rare words, proper nouns, and loanwords.

\textit{Example.} ``She will lead the group to the lead mine.'' The first
\textit{lead} should rhyme with \textit{feed} and the second with \textit{red}.
If the system pronounces both the same way, that is a phonetic accuracy error.
How disruptive it is determines whether you score 1 or 2.

\begin{table}[t]
\small
\centering
\caption{Acoustic channels for each prosodic metric.}
\label{tab:prosodic-channels}
\resizebox{\columnwidth}{!}{%
\begin{tabular}{lcccc}
\toprule
\textbf{Metric} & \textbf{Pitch} & \textbf{Dur.} & \textbf{Loud.} & \textbf{Pause} \\
\midrule \\

Intonation  & \cmark & & & \\
Prosodic Stress     & \cmark & \cmark & \cmark &            \\
Boundary Placement  & \cmark & \cmark &            & \cmark \\
Speech Rate         &            & \cmark &            & \cmark \\
\bottomrule
\end{tabular}%
}
\end{table}

\subsubsection{Lexical Stress}
\label{app:guide_lexical_stress}
\textit{What you are listening for.} Is the stress on the right syllable within
each word? For example, the noun \textit{record} is stressed on the first
syllable (\textsc{re}-cord) while the verb is stressed on the second
(re-\textsc{cord}). This metric is strictly about within-word stress; do not
consider sentence-level emphasis here (see Prosodic Stress, Metric~4).

\textit{Why binary?} Lexical stress errors are relatively unambiguous, and even
a single error can substantially impair word recognition.

\textit{Lexical vs.\ prosodic stress.} Lexical stress is a fixed dictionary
property (which syllable within a word). Prosodic stress (Metric~4) is a
speaker-level choice about which words in a sentence receive prominence.

\paragraph{Prosodic Metrics}

Prosodic metrics ask: is the speech organized and delivered in a way that
communicates the structure and meaning of the sentence? Prosodic errors can
change what a listener understands even when every individual word sounds
correct.

Because pitch, duration, loudness, and pauses serve multiple prosodic
functions, a single acoustic event can be relevant to more than one metric.
Rate each metric for its own question, not the acoustic signal in isolation.
Table~\ref{tab:prosodic-channels} summarizes which channels each metric draws
on.

\subsubsection{Intonation}
\label{app:guide_intonation}
\textit{What you are listening for.} Intonation is the pitch melody across
phrases and sentences. It signals sentence type (questions typically rise;
statements fall), speaker attitude, and discourse structure. This metric
concerns pitch only; do not factor in rate or pauses.

\subsubsection{Prosodic Stress}
\label{app:guide_prosodic_stress}

\textit{What you are listening for.} Which words in the sentence receive extra
prominence? Speakers signal focus via higher pitch, greater loudness, and
longer duration. Do not confuse with lexical stress (Metric~2), which concerns
syllables within a word.

\subsubsection{Prosodic Boundary Placement}
\label{app:guide_prosodic_boundary}

\textit{What you are listening for.} Speakers group words into phrases using
pitch movements, final-syllable lengthening, and pauses. Misplaced boundaries
disrupt parsing even when no alternative meaning is at stake. Focus on whether
the grouping aligns with plausible syntactic structure and produces a pleasant
listening experience.

\textit{Example.} Boundary errors go in both directions. A spurious boundary
(``It was a $\mid$ beautiful day'') sounds choppy; a missing boundary (``When
the dog bites the man screams'') creates a parsing failure. In both cases the
question is whether the boundary made the speech harder or less pleasant to
follow.

\subsubsection{Speech Rate Appropriateness}
\label{app:guide_speech_rate}

\textit{What you are listening for.} Is the overall tempo appropriate, and
does it vary naturally? Human speakers slow for complex or emotionally weighty
material and speed up through predictable or parenthetical content. This metric
is distinct from Boundary Placement
(where phrase breaks occur).

\paragraph*{Paralinguistic Metrics}
Paralinguistic metrics ask: beyond the linguistic message, does the speech
convey appropriate speaker characteristics? This covers emotion,
expressiveness, voice consistency, and whether the audio could have been
produced by a human.

\subsubsection{Emotional Appropriateness}
\label{app:guide_emotional_approrpriateness}
\textit{What you are listening for.} Does the emotional tone of the voice match
the content? A sentence about exciting news should not sound bored, and a
condolence message should not sound cheerful. This metric does not ask whether
the emotion is strong enough (see Expressiveness, Metric~8); it asks only
whether the type of emotion is correct.

\subsubsection{Expressiveness}
\label{app:guide_expressiveness}

\textit{What you are listening for.} Expressiveness captures the degree of
natural variation in delivery. TTS speech can be flat (monotone, robotic) or
overdone (theatrical, exaggerated). Both extremes are errors. Rate
expressiveness relative to what is appropriate for the content: informational
material calls for more measured delivery; emotionally charged content warrants
greater vocal variation.

\subsubsection{Speaker Identity Consistency}
\label{app:guide_speaker_identity}

\textit{What you are listening for.} Does the voice maintain a consistent
perceived identity (stable pitch range, vocal quality, perceived age,
perceived gender, and accent) both within a single sentence and across
sentences in a set? Inconsistencies can occur across sentences or partway
through a single sentence.

\subsubsection{Human Plausibility}
\label{app:guide_human_plausibility}

\textit{What you are listening for.} This metric is independent of linguistic
correctness. The question is whether a human being could have physically
produced this audio. It targets digital artifacts: robotic buzzing, clicks,
metallic resonance, impossible pitch jumps, abruptly cut-off syllables, or
anything outside the range of human vocal production. Do not penalize
linguistic errors here; a mispronounced word is a Phonetic Accuracy issue but
may still sound physically human.

\subsection*{Quick Reference}

\begin{table}[tb]
\small
\centering
\caption{All metrics at a glance. All ternary dimensions are collapsed to binary scoring as described in Appendix \ref{app:human_annotations}.}
\label{tab:metrics-summary}
\resizebox{\columnwidth}{!}{%
\begin{tabular}{llcp{3.0cm}}
\toprule
\textbf{Dim.} & \textbf{Metric} & \textbf{Scale} & \textbf{Summary} \\
\midrule
\multirow{2}{*}{Word} & Phonetic Accuracy       & 1--3 & Right sounds produced? \\
                      & Lexical Stress          & 0/1  & Right syllable stressed? \\
\midrule
\multirow{5}{*}{Pros.} & Intonation             & 1--3 & Pitch melody fits sentence? \\
                       & Prosodic Stress        & 0/1  & Right words emphasized? \\
                       & Boundary Placement     & 1--3 & Natural phrase grouping? \\
                       & Speech Rate            & 1--3 & Tempo appropriate? \\
\midrule
\multirow{4}{*}{Para.} & Emotional Appr.        & 0/1  & Emotion matches content? \\
                       & Expressiveness         & 1--3 & Delivery dynamic/fitting? \\
                       & Speaker Consistency    & 0/1  & Same person throughout? \\
                       & Human Plausibility     & 0/1  & Could a human produce this? \\
\bottomrule
\end{tabular}%
}
\end{table}

\noindent\textbf{Remember:} Rate each metric independently. A single utterance
can score 3 on Phonetic Accuracy but 1 on Intonation, or 0 on Human Plausibility
but 1 on Lexical Stress. Do not let a strong impression from one dimension
bleed into another.

\begin{table*}[t]
\centering
\small
\setlength{\tabcolsep}{5pt}
\caption{%
  \textbf{Intended versus observed label distributions on targeted samples.}
  \textit{Caught} = majority vote agrees with intended failure.
  \textit{Missed} = majority vote labels intended failure as clean.
  \textit{Tie} = no majority among the first 3 valid binary votes
  (only possible when fewer than 3 raters scored the dimension).
  \textit{Miss rate} = missed / intended-fail.
  Bottom panel: of 376 intended-clean items, 363 (96.5\%) were observed clean;
  false-alarm rate = 3.5\%.
}
\label{tab:confusion}
\resizebox{0.90\textwidth}{!}{%
\begin{tabular}{llr r rrr r}
\toprule
\textbf{Tier} & \textbf{Dimension} & \textbf{N}
  & \textbf{Int.\,fail}
  & \textbf{Caught} & \textbf{Missed} & \textbf{Tie}
  & \textbf{Miss rate} \\
\midrule

\multirow{2}{*}{Word-level}
  & Phonetic accuracy              & 120 & 80 & 63 & 17 & 0 & 21.2\% \\
  & Lexical stress                 & 113 & 73 & 45 & 28 & 0 & 38.4\% \\

\midrule

\multirow{4}{*}{Prosodic}
  & Intonation                     & 120 &  80 & 44 & 32 & 4 & 40.0\% \\
  & Prosodic stress                & 124 &  79 & 53 & 26 & 0 & 32.9\% \\
  & Prosodic boundary placement    &  95 &  58 & 45 & 13 & 0 & 22.4\% \\
  & Speech rate                    & 168 & 128 & 93 & 35 & 0 & 27.3\% \\

\midrule

\multirow{4}{*}{Paralinguistic}
  & Emotional appropriateness      &  58 & 39 & 21 & 18 & 0 & 46.2\% \\
  & Expressiveness                 & 100 & 70 & 50 & 20 & 0 & 28.6\% \\
  & Speaker identity consistency   & 104 & 64 & 32 & 31 & 1 & 48.4\% \\
  & Human plausibility             & 115 & 70 & 49 & 19 & 2 & 27.1\% \\

\midrule

  & \textbf{Total}                 & \textbf{1117} & \textbf{741}
    & \textbf{495} & \textbf{239} & \textbf{7} & \textbf{32.3\%} \\

\midrule[0.3pt]

\multicolumn{8}{l}{%
  \textit{Intended-clean items (n\,=\,376):}
  observed clean 363 (96.5\%);
  false-alarm rate 3.5\%;
  ties 0.
}\\

\bottomrule
\end{tabular}%
}
\end{table*}

\subsection{Acoustic Overlap in the Prosodic and Paralinguistic Tiers}
\label{app:acoustic-overlap}

The prosodic dimensions share acoustic channels: f\textsubscript{0}, for instance, carries intonation, prosodic stress, and boundary tones 
simultaneously. Crucially, however, sharing an acoustic channel does not entail these dimensions cannot fail independently. A misplaced prosodic boundary may disrupt the pitch contour without any corresponding failure in accent placement, and a stress error may occur with an otherwise well-formed intonation contour. This independence motivates rating each dimension separately rather than holistically: each stimulus is rated on all applicable dimensions independently, so cases where a single acoustic anomaly produces failures across multiple categories are captured explicitly rather than collapsed into a single quality judgment.

The paralinguistic tier presents an analogous case. These four 
dimensions share acoustic channels with prosody but convey information about the speaker rather than the linguistic message. A synthesis artifact may simultaneously affect perceived speaker identity consistency and human plausibility, yet the two dimensions target distinct speaker properties and can fail independently of one another. 

To assess these overlaps and dependencies, we examined error co-occurrence across dimensions (Figure \ref{fig:correlation_by_dimension}). Error co-occurrence was minor across most dimensions, with several notable exceptions, for which the co-occurrence relationship was unidirectional rather than symmetric: samples with emotional appropriateness errors are frequently rated to also have expressiveness errors; lexical stress errors are often also rated as phonetic accuracy errors; and speaker identity consistency errors are often also rated as human plausibility errors. There were no dimensions for which many error-free samples nonetheless had errors in an unintended dimension, which could have presented as a false positive for a perfectly sensitive TTS evaluator.

\subsection{Paralinguistic Framework and Neural TTS}
\label{app:paralinguistic-framework}

Crystal's original framework distinguishes voice quality features, 
which are speaker-stable properties and remain constant across an utterance, from 
affective paralinguistic features, which modulate in response to 
content and context. The four paralinguistic dimensions in our schema 
map directly onto this distinction: speaker identity consistency and human 
plausibility reflect speaker-stable properties, while expressiveness 
and emotional appropriateness reflect content-driven modulation.

Human plausibility warrants particular note. Modern neural TTS can 
produce voices that are internally consistent and expressively adequate 
but physically implausible (a failure mode not anticipated by 
Crystal's framework but characteristic of neural synthesis). We treat 
it as a distinct paralinguistic dimension rather than subsuming it 
under overall quality, since it represents a novel category of 
synthesis failure with no natural analog in human speech perception 
research.

\subsection{Intended versus observed labels.}
Table~\ref{tab:confusion} reports the agreement between intended quality levels
(design variable) and majority-vote annotator labels (ground truth). Of 741 samples
intended to contain the targeted failure, annotators majority-labeled 239 (32.3\%) as
having no failure: approximately 1 in 3 designed errors was not perceived as such.
Misses are highest for speaker identity consistency
(48.4\%), emotional appropriateness (46.2\%), and intonation (40.0\%), and lowest for prosodic boundary placement
(22.4\%) and phonetic accuracy (21.2\%). The false-alarm rate on intended-clean items
is only 3.5\%, indicating that raters systematically under-call failures rather than
over-call them which is a pattern consistent with the difficulty of eliciting perceptually
salient errors on higher-level dimensions.

\subsection{AUROC Analysis for MOS Models}
\label{sec:auroc_mos}
\begin{table*}[tb]
  \centering
  \footnotesize
  \renewcommand{\arraystretch}{0.85}
  \setlength{\tabcolsep}{3pt}
  \caption{Construct-aligned AUC for MOS models per evaluation dimension.}
  \label{tab:mos_auc}
  \resizebox{\textwidth}{!}{%
  \begin{tabular}{l | ll | llll | llll}
    \toprule
    \textbf{Model} & \multicolumn{2}{c}{\textbf{Word}} & \multicolumn{4}{c}{\textbf{Prosodic}} & \multicolumn{4}{c}{\textbf{Paralinguistic}} \\
    \cmidrule(lr){2-3} \cmidrule(lr){4-7} \cmidrule(lr){8-11}
     & \textbf{Phon.} & \textbf{Lex.} & \textbf{Inton.} & \textbf{Pros.S} & \textbf{Pros.B} & \textbf{Rate} & \textbf{Emo.} & \textbf{Expr.} & \textbf{Spk.} & \textbf{Hum.} \\
    \midrule
    AudioBox$_{\mathrm{CE}}$ & \phantom{-}0.46 & \phantom{-}0.61 & \phantom{-}\textbf{0.76}{\scriptsize$^{***}$} & \phantom{-}\textbf{0.33}{\scriptsize$^{**}$} & \phantom{-}0.47 & \phantom{-}\textbf{0.89}{\scriptsize$^{***}$} & \phantom{-}0.53 & \phantom{-}\textbf{0.63}{\scriptsize$^{*}$} & \phantom{-}\textbf{0.31}{\scriptsize$^{**}$} & \phantom{-}\textbf{0.92}{\scriptsize$^{***}$} \\
    AudioBox$_{\mathrm{CU}}$ & \phantom{-}0.46 & \phantom{-}0.51 & \phantom{-}0.57 & \phantom{-}0.55 & \phantom{-}0.53 & \phantom{-}\textbf{0.84}{\scriptsize$^{***}$} & \phantom{-}0.48 & \phantom{-}\textbf{0.79}{\scriptsize$^{***}$} & \phantom{-}0.38 & \phantom{-}\textbf{0.86}{\scriptsize$^{***}$} \\
    AudioBox$_{\mathrm{PC}}$ & \phantom{-}0.46 & \phantom{-}0.49 & \phantom{-}0.43 & \phantom{-}\textbf{0.74}{\scriptsize$^{***}$} & \phantom{-}0.61 & \phantom{-}0.51 & \phantom{-}0.47 & \phantom{-}\textbf{0.73}{\scriptsize$^{***}$} & \phantom{-}0.40 & \phantom{-}\textbf{0.07}{\scriptsize$^{***}$} \\
    AudioBox$_{\mathrm{PQ}}$ & \phantom{-}0.44 & \phantom{-}0.53 & \phantom{-}\textbf{0.69}{\scriptsize$^{**}$} & \phantom{-}\textbf{0.38}{\scriptsize$^{*}$} & \phantom{-}0.44 & \phantom{-}\textbf{0.81}{\scriptsize$^{***}$} & \phantom{-}0.51 & \phantom{-}\textbf{0.71}{\scriptsize$^{***}$} & \phantom{-}\textbf{0.23}{\scriptsize$^{***}$} & \phantom{-}\textbf{0.92}{\scriptsize$^{***}$} \\
    DNSMOS-Pro$_{\mathrm{BVCC}}$ & \phantom{-}0.42 & \phantom{-}0.46 & \phantom{-}0.56 & \phantom{-}0.59 & \phantom{-}0.63 & \phantom{-}\textbf{0.40}{\scriptsize$^{*}$} & \phantom{-}0.46 & \phantom{-}0.45 & \phantom{-}0.36 & \phantom{-}\textbf{0.82}{\scriptsize$^{***}$} \\
    DNSMOS-Pro$_{\mathrm{VCC}}$ & \phantom{-}0.50 & \phantom{-}0.41 & \phantom{-}0.52 & \phantom{-}0.47 & \phantom{-}0.47 & \phantom{-}0.55 & \phantom{-}0.59 & \phantom{-}\textbf{0.79}{\scriptsize$^{***}$} & \phantom{-}0.44 & \phantom{-}\textbf{0.78}{\scriptsize$^{***}$} \\
    NISQA & \phantom{-}0.52 & \phantom{-}0.47 & \phantom{-}0.55 & \phantom{-}0.60 & \phantom{-}0.57 & \phantom{-}\textbf{0.67}{\scriptsize$^{**}$} & \phantom{-}0.49 & \phantom{-}\textbf{0.74}{\scriptsize$^{***}$} & \phantom{-}0.47 & \phantom{-}\textbf{0.76}{\scriptsize$^{***}$} \\
    UTMOSv2 & \phantom{-}0.50 & \phantom{-}0.56 & \phantom{-}\textbf{0.70}{\scriptsize$^{**}$} & \phantom{-}\textbf{0.64}{\scriptsize$^{*}$} & \phantom{-}0.48 & \phantom{-}\textbf{0.77}{\scriptsize$^{***}$} & \phantom{-}0.46 & \phantom{-}0.57 & \phantom{-}0.53 & \phantom{-}\textbf{0.72}{\scriptsize$^{**}$} \\
    \bottomrule
  \end{tabular}%
  }

  \vspace{4pt}
  \parbox{\textwidth}{\footnotesize
    AudioBox subscripts: CE\,=\,Content Enjoyment; CU\,=\,Content Usefulness; PC\,=\,Production Complexity; PQ\,=\,Production Quality. DNSMOS-Pro subscripts: BVCC\,=\,trained on BVCC corpus; VCC\,=\,trained on VCC2018 corpus.
    \textbf{Column key:} Phon.\,=\,Phonetic Accuracy; Lex.\,=\,Lexical Stress; Inton.\,=\,Intonation; Pros.S\,=\,Prosodic Stress; Pros.B\,=\,Prosodic Boundary; Rate\,=\,Speech Rate; Emo.\,=\,Emotional Appropriateness; Expr.\,=\,Expressiveness; Spk.\,=\,Speaker Identity; Hum.\,=\,Human Plausibility.
    \textbf{Bold} = significant ($p < .05$). $^{*}p<.05$, $^{**}p<.01$, $^{***}p<.001$ (two-sided Mann--Whitney $U$).}
\end{table*}
\begin{table*}[htbp]
  \centering
  \footnotesize
  \renewcommand{\arraystretch}{0.85}
  \setlength{\tabcolsep}{3pt}
  \caption{Per-dimension Kendall's $\tau$ against human ground truth on the unanimous subset ($N=516$). \textsc{\audiollm} judge rows use the C1 naturalness prompt}
  \label{tab:unanimous-robustness}
  \resizebox{\textwidth}{!}{%
  \begin{tabular}{l | ll | llll | llll}
    \toprule
    \textbf{Model} & \multicolumn{2}{c}{\textbf{Word}} & \multicolumn{4}{c}{\textbf{Prosodic}} & \multicolumn{4}{c}{\textbf{Paralinguistic}} \\
    \cmidrule(lr){2-3} \cmidrule(lr){4-7} \cmidrule(lr){8-11}
     & \textbf{Phon.} & \textbf{Lex.} & \textbf{Inton.} & \textbf{Pros.S} & \textbf{Pros.B} & \textbf{Rate} & \textbf{Emo.} & \textbf{Expr.} & \textbf{Spk.} & \textbf{Hum.} \\
    \midrule
    \midrule
    \multicolumn{11}{c}{MOS Models} \\
    \midrule
    AudioBox$_{\mathrm{CE}}$ & -0.066 & \phantom{-}0.154 & \phantom{-}\textbf{0.338}{\scriptsize$^{*}$} & \textbf{-0.246}{\scriptsize$^{*}$} & -0.029 & \phantom{-}\textbf{0.537}{\scriptsize$^{***}$} & \phantom{-}0.064 & \phantom{-}\textbf{0.308}{\scriptsize$^{*}$} & \textbf{-0.459}{\scriptsize$^{***}$} & \phantom{-}\textbf{0.651}{\scriptsize$^{***}$} \\
    AudioBox$_{\mathrm{CU}}$ & -0.058 & \phantom{-}0.018 & \phantom{-}0.059 & \phantom{-}0.047 & -0.122 & \phantom{-}\textbf{0.458}{\scriptsize$^{***}$} & \phantom{-}0.081 & \phantom{-}\textbf{0.673}{\scriptsize$^{***}$} & \textbf{-0.320}{\scriptsize$^{*}$} & \phantom{-}\textbf{0.584}{\scriptsize$^{***}$} \\
    AudioBox$_{\mathrm{PC}}$ & -0.014 & \phantom{-}0.020 & -0.022 & \phantom{-}\textbf{0.318}{\scriptsize$^{**}$} & \phantom{-}0.062 & -0.003 & -0.226 & \phantom{-}\textbf{0.359}{\scriptsize$^{**}$} & -0.066 & \textbf{-0.629}{\scriptsize$^{***}$} \\
    AudioBox$_{\mathrm{PQ}}$ & -0.097 & \phantom{-}0.071 & \phantom{-}0.250 & -0.096 & \textbf{-0.243}{\scriptsize$^{*}$} & \phantom{-}\textbf{0.428}{\scriptsize$^{***}$} & \phantom{-}0.047 & \phantom{-}\textbf{0.582}{\scriptsize$^{***}$} & \textbf{-0.546}{\scriptsize$^{***}$} & \phantom{-}\textbf{0.645}{\scriptsize$^{***}$} \\
    DNSMOS-Pro$_{\mathrm{BVCC}}$ & -0.105 & -0.088 & \phantom{-}0.044 & \phantom{-}0.143 & \phantom{-}0.119 & -0.126 & -0.132 & -0.174 & \textbf{-0.332}{\scriptsize$^{*}$} & \phantom{-}\textbf{0.410}{\scriptsize$^{**}$} \\
    DNSMOS-Pro$_{\mathrm{VCC}}$ & \phantom{-}0.022 & -0.101 & \phantom{-}0.044 & -0.016 & -0.027 & \phantom{-}0.045 & \phantom{-}0.206 & \phantom{-}\textbf{0.534}{\scriptsize$^{***}$} & -0.054 & \phantom{-}\textbf{0.460}{\scriptsize$^{**}$} \\
    NISQA & \phantom{-}0.060 & -0.033 & \phantom{-}0.294 & \phantom{-}0.171 & \phantom{-}0.066 & \phantom{-}\textbf{0.253}{\scriptsize$^{**}$} & \phantom{-}0.073 & \phantom{-}\textbf{0.637}{\scriptsize$^{***}$} & -0.109 & \phantom{-}\textbf{0.455}{\scriptsize$^{**}$} \\
    UTMOSv2 & \phantom{-}0.040 & \phantom{-}0.077 & \phantom{-}0.276 & \phantom{-}0.125 & \phantom{-}0.124 & \phantom{-}\textbf{0.465}{\scriptsize$^{***}$} & \phantom{-}0.077 & \phantom{-}0.158 & \phantom{-}0.070 & \phantom{-}\textbf{0.427}{\scriptsize$^{**}$} \\
    \midrule
    \multicolumn{11}{c}{AudioLLM Models} \\
    \midrule
    Gemini 3 Flash$^\dag$ & \phantom{-}\textbf{0.263}{\scriptsize$^{*}$} & \phantom{-}\textbf{0.277}{\scriptsize$^{*}$} & \phantom{-}0.358 & \phantom{-}0.067 & \phantom{-}0.170 & \phantom{-}0.214 & \phantom{-}\textbf{0.433}{\scriptsize$^{*}$} & \phantom{-}\textbf{0.315}{\scriptsize$^{*}$} & -0.062 & \phantom{-}0.289 \\
    Gemini 3 Flash$^\triangle$ & -0.067 & \phantom{-}0.000 & -0.219 & \phantom{-}0.154 & \phantom{-}0.119 & -0.029 & -0.082 & -0.051 & -0.078 & \phantom{-}0.227 \\
    Gemini 3.5 Flash$^\dag$ & \phantom{-}\textbf{0.376}{\scriptsize$^{**}$} & \phantom{-}\textbf{0.296}{\scriptsize$^{*}$} & \phantom{-}0.280 & \phantom{-}0.225 & \phantom{-}\textbf{0.380}{\scriptsize$^{**}$} & \phantom{-}0.082 & \phantom{-}\textbf{0.412}{\scriptsize$^{*}$} & \phantom{-}\textbf{0.297}{\scriptsize$^{*}$} & \phantom{-}0.258 & \phantom{-}0.322 \\
    Gemini 3.5 Flash$^\triangle$ & -0.103 & \phantom{-}0.236 & -0.144 & \phantom{-}0.188 & \phantom{-}\textbf{0.427}{\scriptsize$^{***}$} & \phantom{-}0.028 & \phantom{-}0.298 & \phantom{-}0.139 & \phantom{-}0.062 & \phantom{-}\textbf{0.378}{\scriptsize$^{*}$} \\
    Qwen3 Omni$^\dag$ & \phantom{-}0.000 & -0.057 & . & -0.131 & \phantom{-}\textbf{0.255}{\scriptsize$^{*}$} & \phantom{-}0.187 & \phantom{-}0.000 & \phantom{-}0.149 & \phantom{-}\textbf{0.388}{\scriptsize$^{*}$} & \phantom{-}0.107 \\
    Qwen3 Omni$^\triangle$ & \phantom{-}0.088 & \phantom{-}0.223 & -0.076 & -0.067 & \phantom{-}\textbf{0.258}{\scriptsize$^{*}$} & \phantom{-}\textbf{0.299}{\scriptsize$^{**}$} & \phantom{-}0.289 & \phantom{-}\textbf{0.357}{\scriptsize$^{*}$} & \phantom{-}0.218 & \phantom{-}0.270 \\
    Step Audio 2 Mini$^\dag$ & \phantom{-}0.000 & -0.031 & -0.158 & \phantom{-}0.000 & \phantom{-}0.147 & -0.077 & -0.215 & \phantom{-}0.189 & \phantom{-}0.226 & \phantom{-}0.066 \\
    Step Audio 2 Mini$^\triangle$ & \phantom{-}0.115 & \phantom{-}0.033 & \phantom{-}0.072 & \phantom{-}0.097 & -0.246 & -0.098 & -0.158 & \phantom{-}0.183 & \phantom{-}0.056 & \phantom{-}0.188 \\
    \bottomrule
  \end{tabular}%
  }

  \vspace{4pt}
  \parbox{\textwidth}{\footnotesize
    \textbf{Model Subscripts:} CE\,=\,Content Enjoyment; CU\,=\,Content Usefulness; PC\,=\,Production Complexity; PQ\,=\,Production Quality. DNSMOS-Pro subscripts: BVCC\,=\,trained on BVCC corpus; VCC\,=\,trained on VCC2018 corpus.
    The unanimous subset contains 516 items for which all three annotators agreed on the matched rating dimension.
    \textsc{AudioLLM} rows use the C1 naturalness prompt.
    \textbf{Columns:} Phon.\,=\,Phonetic Accuracy; Lex.\,=\,Lexical Stress; Inton.\,=\,Intonation; Pros.S\,=\,Prosodic Stress; Pros.B\,=\,Prosodic Boundary; Rate\,=\,Speech Rate; Emo.\,=\,Emotional Appropriateness; Expr.\,=\,Expressiveness; Spk.\,=\,Speaker Identity; Hum.\,=\,Human Plausibility.
    \textbf{Transcript:}
    $^\dag$\,=\,with transcript; $^\triangle$\,=\,without transcript.
    \textbf{Bold} = significant ($p < .05$). $^{*}p<.05$, $^{**}p<.01$, $^{***}p<.001$ (two-sided).
    A dot indicates undefined $\tau$ because the model produced a constant prediction.}
\end{table*}

As demonstrated in Table \ref{tab:mos_auc}, MOS models exhibit striking performance gap across dimensions. Most presented models achieve strong alignment with human judgments on human plausibility (AUC $0.72-0.92$) and speech rate (AUC $0.40-0.89$), yet systematically fail on word-level and lower prosodic dimensions. Phonetic accuracy and lexical stress yield near-chance performance across all models (AUC $0.41-0.61$), while intonation and prosodic stress show modest, inconsistent correlation (AUC $0.33-0.76$).

\subsection{Prompt Details}
\label{sec:prompt_details}
\begin{tcolorbox}[promptbox, title={\small Condition 1: Underspecified MOS-style prompt}]
       [ROLE]
        You are a speech quality evaluator assessing text-to-speech audio. \\

        [TASK]
        {\% if show\_text \%}A synthesized speech sample is attached above, along with the text it was
        generated from. Evaluate the overall naturalness of the speech.

        Intended text: "{{original\_text}}"{\% else \%}A synthesized speech sample is attached above. Evaluate its overall
        naturalness based on the audio alone.{\% endif \%} \\

        [CRITERIA]
        Rate the naturalness on a binary scale (0 = failure present, 1 = no failure):
          0: One or more issues affecting naturalness are perceivable, whether minor or severe.
          1: Completely natural, indistinguishable from human speech. \\
        
        [PROCEDURE]
        Follow these steps:
        1. Attend to the attached audio sample.
        {\% if show\_text \%}2. Compare the spoken output to the intended text.
        3. Identify any issues that affect naturalness.
        4. Decide the most appropriate score.{\% else \%}2. Identify any issues that affect naturalness.
        3. Decide the most appropriate score.{\% endif \%}\\
        
        [OUTPUT]
        Respond in this exact JSON format:
        {
          "evaluation\_steps": "<your step-by-step reasoning>",
          "score": <0 or 1>
        }
\end{tcolorbox}

\begin{tcolorbox}[promptbox, title={\small Condition 2a: Schema-guided prompt, single}]

[ROLE]
        You are an expert speech quality evaluator with training in
        phonetics and prosodic phonology. \\

        [TASK]
        {\% if show\_text \%}A synthesized speech sample is attached above, along with the text it was
        generated from. Evaluate the overall quality of the speech,
        considering all dimensions in the evaluation schema below.

        Intended text: "{{original\_text}}"{\% else \%}A synthesized speech sample is attached above. Evaluate the overall
        quality of the speech, considering all dimensions in the evaluation
        schema below.{\% endif \%} \\

        [CRITERIA]
        Consider the following ten dimensions when forming your
        overall judgment: \\

        WORD-LEVEL:
        - Phonetic Accuracy: Does every word contain the right speech
          sounds? This includes consonants, vowels, and the way adjacent
          sounds blend together (coarticulation). Check whether the system
          produced the correct phonemes for each word, especially for
          heteronyms, rare words, proper nouns, and loanwords.
        - Lexical Stress: Is the stress on the right syllable within each
          word? Every multi-syllable English word has a predictable stress
          pattern. Incorrect lexical stress often changes vowel quality
          too. This is strictly about within-word stress. \\

        PROSODIC:
        - Intonation: The melody of speech: the way pitch rises and falls
          across phrases and sentences. It signals sentence type, speaker
          attitude, and discourse structure. Concerns pitch only.
        - Prosodic Stress: Which words in a sentence receive extra
          prominence via higher pitch, greater loudness, and longer
          duration, signaling which information is new, important, or
          contrastive. This is about entire words within a sentence,
          not syllables within a word.
        - Prosodic Boundary Placement: Speakers group words into phrases
          using pitch movements, slight lengthening of the final syllable,
          and pauses. Misplaced boundaries disrupt the listening
          experience. Focus on whether grouping aligns with plausible
          syntactic structure.
        - Speech Rate Appropriateness: Is the overall speed appropriate,
          and does it vary in natural ways? Human speakers slow down for
          complex material and speed up through predictable material. \\

        PARALINGUISTIC:
        - Emotional Appropriateness: Does the emotional tone of the voice
          match the content of the text? This does not ask whether the
          emotion is strong enough (that is Expressiveness). It asks only
          whether the type of emotion is correct.
        - Expressiveness: Natural variation in delivery style, pace, and
          vocal energy. Human speech is dynamic. TTS speech can be flat
          or overdone. Rate relative to what would be appropriate for the
          content.
        - Speaker Identity Consistency: Does the voice maintain a
          consistent perceived identity in terms of pitch range, vocal
          quality, perceived age, gender, and accent? Check for shifts
          both within and across sentences.
        - Human Plausibility: Could a human have physically produced this
          audio? Targets digital artifacts and glitches: robotic buzzing,
          clicks, unnatural metallic resonance, impossible pitch jumps,
          abrupt syllable cutoffs. Independent of linguistic correctness. \\

        Considering ALL of the above, rate overall quality on a binary scale
        (0 = failure present, 1 = no failure):
          0: One or more issues detected across the ten dimensions, whether minor or severe.
          1: No perceivable issues detected across all ten dimensions. \\

        [PROCEDURE]
        Follow these steps:
        1. Attend to the attached audio sample.
        2. For each of the three dimension groups (word-level, prosodic,
           paralinguistic), note whether any issues are present.
        3. Identify which specific dimensions (if any) show failures.
        4. Decide the most appropriate score. \\

        [OUTPUT]
        Respond in this exact JSON format:
        {
          "word\_level\_issues": "<any word-level issues noted>",
          "prosodic\_issues": "<any prosodic issues noted>",
          "paralinguistic\_issues": "<any paralinguistic issues noted>",
          "score": <0 or 1>
        }
\end{tcolorbox}

\begin{tcolorbox}[promptbox, title={\small Condition 2b: Schema-guided prompt, per-dimension scores}]

      [ROLE]
        You are an expert speech quality evaluator with training in
        phonetics and prosodic phonology. \\

        [TASK]
        {\% if show\_text \%}A synthesized speech sample is attached above, along with the text it was
        generated from. Evaluate the speech independently on each of the
        ten dimensions below, using the specified scale for each. \\

        Intended text: "{{original\_text}}"{\% else \%}A synthesized speech sample is attached above. Evaluate the speech
        independently on each of the ten dimensions below, using the
        specified scale for each.{\% endif \%} \\

        [CRITERIA]
        Rate each metric independently. Do not let a strong impression
        from one dimension bleed into another. \\

        ALL DIMENSIONS use a binary scale (0 = failure present, 1 = no failure):

        Lexical Stress:
        Is the stress on the right syllable within each word? This is
        strictly about within-word stress, not sentence-level emphasis.
          0: At least one word has stress on the wrong syllable.
          1: Every multi-syllable word has stress on the correct syllable. \\

        Prosodic Stress:
        Which words in the sentence receive extra prominence via higher
        pitch, greater loudness, and longer duration?
          0: Prominence falls on words that do not warrant it, or words
             that should be prominent are not.
          1: The choice of which words receive prominence sounds
             predictable and appropriate. \\

        Emotional Appropriateness:
        Does the emotion of the voice match the content of the text? If the text is neutral, the emotion should be neutral as well.
          0: The emotion is wrong for the content.
          1: The emotion is a plausible match for the text. \\

        Speaker Identity Consistency:
        Does the voice maintain a consistent perceived identity?
          0: The voice changes in a way that makes it sound like a
             different person.
          1: The voice is recognizably consistent throughout. \\

        Human Plausibility:
        Could a human have physically produced this audio?
          0: The audio contains at least one moment that could not have
             been produced by a human speaker.
          1: The entire audio stream sounds like it could have been
             produced by a human speaker. \\

        Phonetic Accuracy:
          0: One or more consonants or vowels are outside the range of possible expected sounds for the utterance spoken in general american english.
          1: All consonants and vowels fall within the range of possible expected sounds for the utterance spoken in general american english.

        Intonation:
          0: Perceivable errors in the pitch contour, regardless of severity.
          1: No perceivable errors in the pitch contour. \\

        Prosodic Boundary Placement:
          0: Misplaced boundary disrupts sentence structure, noticeably or severely.
          1: Phrase grouping aligns naturally with syntactic structure. \\

        Speech Rate Appropriateness:
          0: Speech rate is too slow or too fast during parts or the whole of the utterance.
          1: Speech rate feels natural and varies appropriately. \\

        Expressiveness: \\
          0: Delivery is too flat or too dramatic, regardless of severity.
          1: Delivery has natural, appropriate variation. \\

        [PROCEDURE]
        For each dimension:
        1. Attend to the attached audio sample focusing on that dimension.
        2. Compare what you hear to the rubric anchors above.
        3. Record your reasoning and score.

        Evaluate each dimension independently. \\

        [OUTPUT]
        Respond in this exact JSON format:
\begin{lstlisting}[
    style=jsonblock,
    basicstyle=\scriptsize\ttfamily,
    columns=flexible,
    breaklines=true
]
{
  "phonetic_accuracy":{"reasoning": "...", "score": <0 or 1>},
  "lexical_stress":{"reasoning": "...", "score": <0 or 1>},
  "intonation": {"reasoning": "...", "score": <0 or 1>},
  "prosodic_stress":{"reasoning": "...", "score": <0 or 1>},
  "prosodic_boundary":{"reasoning": "...", "score": <0 or 1>},
  "speech_rate": {"reasoning": "...", "score": <0 or 1>},
  "emotional_appropriateness":  {"reasoning": "...", "score": <0 or 1>},
  "expressiveness":{"reasoning": "...", "score": <0 or 1>},
  "speaker_identity":{"reasoning": "...", "score": <0 or 1>},
  "human_plausibility":{"reasoning": "...", "score": <0 or 1>}
}
\end{lstlisting}

\end{tcolorbox}

\begin{tcolorbox}[promptbox, title={\small Condition 3: Isolated per-dimension prompts}]

[ROLE]
        You are an expert speech quality evaluator with training in
        phonetics and prosodic phonology. \\

        [TASK]
        {\% if show\_text \%}A synthesized speech sample is attached above, along with the text it was
        generated from. Evaluate the speech on ONE specific dimension.

        Intended text: "{{original\_text}}"{\% else \%}A synthesized speech sample is attached above. Evaluate the speech
        on ONE specific dimension based on the audio alone.{\% endif \%} \\

        [CRITERIA]
        {{dimension\_block}} \\

        [PROCEDURE]
        Follow these steps:
        1. Attend to the attached audio sample.
        2. Focus exclusively on {{dimension\_name}}.
        3. Compare what you hear to the score anchors above.
        4. Select your score.

        Do not consider any other aspect of speech quality. Evaluate
        ONLY {{dimension\_name}}.

        [OUTPUT]
        Respond in this exact JSON format:
        {
          "reasoning": "<your reasoning about {{dimension\_name}}>",
          "score": <{{score\_range}}>
        }
\end{tcolorbox}


\section{Generation Pipeline Details}
\label{app:pipeline}

This appendix documents the model configurations and
hyperparameters for the error generation pipeline
described in Section~\ref{sec:error-gen-pipeline}.
All audio: 44,100\,Hz, 16-bit PCM, mono.
All dimensions use binary labels: 1 (acceptable)
and 0 (error). For dimensions originally generated
with three severity levels, scores 1 and 2 were
collapsed to label 0; score 3 maps to label 1.

\subsection{Shared Defaults}
\label{app:common_params}

\noindent\textbf{LLM.} Azure OpenAI \texttt{gpt-5};
\texttt{api\_version}
\texttt{2025-03-01-preview};
\texttt{max\_completion\_tokens} = 16384;
temperature: API default.

\noindent\textbf{TTS (default).}
Cartesia \texttt{sonic-3}:
\begin{itemize}[leftmargin=*, nosep]
    \item Blake: \texttt{\seqsplit{a167e0f3-df7e-4d52-a9c3-f949145efdab}}
    \item Brooke: \texttt{\seqsplit{e07c00bc-4134-4eae-9ea4-1a55fb45746b}}
    \item Emma: \texttt{\seqsplit{f6ff7c0c-e396-40a9-a70b-f7607edb6937}}
    \item Ronald: \texttt{\seqsplit{5ee9feff-1265-424a-9d7f-8e4d431a12c7}}
    \item Henry: \texttt{\seqsplit{87286a8d-7ea7-4235-a41a-dd9fa6630feb}}
\end{itemize}
\noindent\textbf{TTS (intonation).}
ElevenLabs \texttt{eleven\_multilingual\_v2}.

\subsection{Pathway 1: IPA-Controlled TTS}
\label{app:ipa_tts}

\texttt{gpt-5} produces a modified transcript for
each intended error level, passed to Cartesia
Sonic-3.
Table~\ref{tab:ipa_voices} lists voices and prompts are listed in \ref{app:llm_prompts};
Table~\ref{tab:ipa_scores} lists intended errors.

\begin{table}[t]
\centering
\small
\caption{IPA pathway: voice per
dimension. All use Cartesia Sonic-3.}
\label{tab:ipa_voices}
\begin{tabular}{lll}
\toprule
\textbf{Dim.} & \textbf{Voice} \\
\midrule
Phon.\ Acc. & Brooke  \\
Lex.\ Str.  & Blake       \\
Pros.\ Str. & Blake       \\
\bottomrule
\end{tabular}
\end{table}

\begin{table}[t]
\centering
\small
\caption{IPA pathway: intended error per label.}
\label{tab:ipa_scores}
\begin{tabular}{llp{3.5cm}}
\toprule
\textbf{Dim.} & \textbf{Label} &
  \textbf{Intended error} \\
\midrule
Phon. & 1 & Canonical IPA \\
Acc.  & 0 & Phoneme substitution
             (within- or cross-class) \\
\midrule
Lex.  & 1 & Canonical stress \\
Str.  & 0 & Shifted stress +
             vowel reduction \\
\midrule
Pros. & 1 & Emphasis on content word \\
Str.  & 0 & Emphasis on function word
             (Cartesia quote-markup) \\
\bottomrule
\end{tabular}
\end{table}

\subsection{Pathway 2: Acoustic Manipulation}
\label{app:praat}

A clean TTS baseline is synthesized first;
Parselmouth/Praat PSOLA or local DSP then
introduces a targeted perturbation.

\paragraph{Intonation.}
Source: EmergentTTS-Questions split (declaratives
and questions; 4--12 words; no commas/dashes).
ElevenLabs voice Brian
(\texttt{nPczCjzI2dev\-NBz1zQrb})
serves as the label-1 anchor; label-0 samples are
derived via F0 manipulation with the Praat
parameters in Table~\ref{tab:praat_params}.

\begin{table}[t]
\centering
\small
\caption{Praat PSOLA parameters (Intonation
and Speaker Identity).}
\label{tab:praat_params}
\begin{tabular}{lr}
\toprule
\textbf{Parameter} & \textbf{Value} \\
\midrule
Analysis floor   & 75\,Hz \\
Manipulation floor & 50\,Hz \\
Pitch ceiling    & 600\,Hz \\
Output clip range & 60--320\,Hz \\
Control points   & 45 \\
Time step        & 0.01\,s \\
\bottomrule
\end{tabular}
\end{table}

\noindent\textbf{Label 1:} verbatim ElevenLabs
synthesis.
\textbf{Label 0:} F0 contour manipulation ;
either terminal-rise compression (last
${\sim}$30\% of contour compressed to 15--30\%
of original excursion) or one of 10 high-impact
strategies (Table~\ref{tab:f0_strategies}),
randomly selected. Two strategies may be stacked
with probability \texttt{--stack-rate} $P$
(default 0.0).

\begin{table}[t]
\small
\caption{F0 strategies for Intonation label 0.
All operate within mean $\pm$ 70--130\,Hz.}
\label{tab:f0_strategies}
\begin{tabular}{lp{3.4cm}c}
\toprule
\textbf{Strategy} & \textbf{Parameters} & \textbf{Kept} \\
\midrule
terminal-rise comp. & Last ${\sim}$30\% of contour compressed to 15--30\% of original excursion & \checkmark \\
reverse        & Reverse contour; scale 1.2--1.5$\times$; $\pm$25\,Hz noise & \checkmark \\
invert         & Invert around mean; scale 1.3--1.6$\times$; $\pm$25\,Hz noise & \checkmark \\
question       & Linear rise: mean$-$40 $\to$ mean$+$130\,Hz & \checkmark \\
exaggerate     & Range $\times$3.5--5.0; clip at ceiling & \checkmark \\
staircase      & 3--5 levels; mean $\pm$70\,Hz & \checkmark \\
random jumps   & ${\sim}$50\% frames random; mean $\pm$90\,Hz & \checkmark \\
flatten+spike  & 0.7$\times$mean base; 3--5 bumps 0.5--1.6$\times$ & --- \\
sine wave      & 3--5\,Hz wobble; $\pm$40--65\,Hz & \checkmark \\
monotone drift & 130$\to$180\,Hz drift; $\pm$30\,Hz wobble & --- \\
random walk    & Cumulative; $\sigma\approx$35\,Hz/step & \checkmark \\
\bottomrule
\end{tabular}
\end{table}

\paragraph{Speech Rate.}
Voice: default English. Direction
(\textit{fast}/\textit{slow}) randomly assigned.
Label 1: Cartesia \texttt{speed=1.0}.
Label 0: either Cartesia \texttt{speed=1.4}
(fast) or \texttt{0.6} (slow), optionally
further stretched via \texttt{librosa}
time-stretch --- factor 0.7 (fast; net
${\approx}$2.0$\times$) or 1.3 (slow; net
${\approx}$0.78$\times$).

\paragraph{Speaker Identity Consistency.}
Each sample is a pair: label 0 (inconsistent)
and label 1 (consistent).
Label 0 uses either (a) equal-power crossfade
of two renderings by different voices across a
target vowel ($\geq$50\,ms), or (b) stitch of
two utterances with a 400\,ms gap.
Voice B undergoes PSOLA pitch alignment when
F0 mismatch exceeds 0.5 semitones; pairs with
$>$3 semitones are rejected.
Voice pairs: Brooke $\leftrightarrow$ Emma
(feminine), Ronald $\leftrightarrow$ Henry
(masculine), all Cartesia Sonic-3 SSE.

\paragraph{Human Plausibility.}
\texttt{gpt-5} selects 1--2 artifact types from
Table~\ref{tab:artifact_taxonomy} (always from
distinct categories); local DSP applies the
recipe to clean Cartesia synthesis (default
English voice).
Label 1: unmodified; label 0: post-processed.

\begin{table}[t]
\caption{DSP artifact taxonomy
(25 types, 8 categories).}
\label{tab:artifact_taxonomy}
\begin{tabular}{lp{4.6cm}}
\toprule
\textbf{Category} & \textbf{Types (\checkmark{} = used in kept dataset)} \\
\midrule
Compression & mp3 compression \checkmark, quantization noise \checkmark \\
Clipping    & hard clipping \checkmark, soft clipping \checkmark, bit crushing \checkmark \\
Glitch      & buffer dropout, stutter repeat , granular fragmentation \checkmark \\
Noise       & white noise \checkmark, pink noise, vinyl crackle \checkmark, EM hum 50\,Hz \checkmark, EM hum 60\,Hz \checkmark \\
Time/Pitch  & time-stretch artifact \checkmark, pitch-correction artifact \checkmark, phase cancellation \checkmark \\
Digital     & bit errors \checkmark, sample-rate mismatch, buffer underrun \checkmark \\
Delay       & echo feedback \checkmark, resonant comb filter \checkmark, feedback squeal \checkmark \\
Spectral    & FFT smearing \checkmark, phase distortion \checkmark, aliasing \checkmark \\
\bottomrule
\end{tabular}
\end{table}

\subsection{Pathway 3: Synthetically Generated Text}
\label{app:synthetic_text}

\paragraph{Prosodic Boundary Placement.}
We created 60 manually curated sentences from Claude Opus 4.8, stratified by error severity and type. The set contains 20 sentences with severe errors (Score: 0), 20 with moderate errors (Score: 0), and 20 with correct boundaries (Score: 1). Within each tier, 10 sentences contain boundary insertion errors and 10 contain boundary deletion errors.

The severe-deletion subset uses garden-path sentences (e.g., ``the old man the boat'') where a necessary boundary deletion forces readers to reparse syntactic structure. This is a failure mode where both human readers and modern TTS systems consistently underperform. Severe-insertion sentences feature mid-constituent commas that create infelicitous boundaries (e.g., ``The quick brown, fox jumps over the lazy dog''). 

Moderate-deletion errors are produced by deleting spaces around coordinating conjunctions, creating rushed but recoverable runs (e.g., ``The rain stoppedso we went for a walk''). Moderate-insertion errors place commas between major constituents that listeners can still regroup (e.g., ``He bought, a new car last weekend'').
TTS: Cartesia Sonic-3; LLM-altered plain text pipeline.

\paragraph{Emotional Appropriateness.}
38 items use Cartesia Sonic-3 with a mismatched
inline emotion tag (e.g.\ \texttt{euphoric}
on sad text) for label 0.
20 additional items use ElevenLabs voice Gerald
(\textit{Emotionless, Flat and Clear}) for flat
delivery of emotionally loaded sentences (label 0).
Label 1: matched emotion or natural delivery.

\begin{table}[t]
\centering
\small
\caption{TTS APIs and source datasets used.}
\label{tab:artifacts_apis}
\resizebox{\columnwidth}{!}{%

\begin{tabular}{p{3.0cm} p{3.0cm} p{1.5cm}}
\toprule
\textbf{Artifact} & \textbf{Version / Model} & \textbf{License} \\
\midrule
\multicolumn{3}{l}{\textit{TTS APIs / Services}} \\
\midrule
Cartesia TTS API & cartesia SDK & Commercial \\
ElevenLabs TTS API & eleven\_multilingual\_v2 & Commercial \\
Azure OpenAI / GPT-5 & openai SDK & Commercial \\
\midrule
\multicolumn{3}{l}{\textit{Source Datasets}} \\
\midrule
Harvard Sentences & IEEE Std 269 & IEEE \\
EmergentTTS-Eval & bosonai/EmergentTTS-Eval & Apache 2.0 \\
\bottomrule
\end{tabular}%
}
\end{table}

\begin{table}[t]
\centering
\small
\caption{Signal processing and audio tools used.}
\label{tab:artifacts_audio}
\resizebox{\columnwidth}{!}{%
\begin{tabular}{p{3.5cm} p{2.5cm} p{1.5cm}}
\toprule
\textbf{Artifact} & \textbf{Version / Notes} & \textbf{License} \\
\midrule
\multicolumn{3}{l}{\textit{Signal Processing / Audio Tools}} \\
\midrule
Praat (praat-parselmouth) & parselmouth & GPL v3 \\
Montreal Forced Aligner & english\_us\_arpa & MIT \\
SciPy & signal processing & BSD 3-Clause \\
SoundFile / libsndfile & audio I/O & BSD 3-Clause \\
\bottomrule
\end{tabular}%
}
\end{table}

\begin{table*}[t]
\centering
\small
\caption{Python packages used in the pipeline and analysis stages of this work.}
\label{tab:artifacts_packages}
\resizebox{0.8\textwidth}{!}{%
\begin{tabular}{p{3.5cm} p{5.0cm} p{2.5cm}}
\toprule
\textbf{Package} & \textbf{Purpose} & \textbf{License} \\
\midrule
\multicolumn{3}{l}{\textit{Pipeline Packages}} \\
\midrule
openai            & Azure OpenAI / GPT-5 API client      & MIT \\
cartesia          & Cartesia TTS API client               & MIT \\
elevenlabs        & ElevenLabs TTS API client             & MIT \\
numpy             & Numerical computation                 & BSD 3-Clause \\
pandas            & Tabular data manipulation             & BSD 3-Clause \\
datasets          & HuggingFace dataset loading           & Apache 2.0 \\
huggingface\_hub  & HuggingFace model/data access         & Apache 2.0 \\
openpyxl          & Excel file I/O                        & MIT \\
python-dotenv     & Environment variable management       & BSD 3-Clause \\
ruamel-yaml       & YAML configuration parsing            & MIT \\
soundfile         & Audio file I/O                        & BSD 3-Clause \\
\midrule
\multicolumn{3}{l}{\textit{Analysis Packages}} \\
\midrule
krippendorff      & Inter-annotator agreement             & MIT \\
scikit-learn      & Classification and evaluation metrics & BSD 3-Clause \\
statsmodels       & Statistical modeling                  & BSD 3-Clause \\
pingouin          & Statistical tests                     & BSD 3-Clause \\
crowd-kit         & Crowdsourcing aggregation             & Apache 2.0 \\
dominance-analysis & Relative importance analysis         & MIT \\
plotly            & Interactive visualization             & MIT \\
matplotlib        & Static visualization                  & PSF / BSD \\
seaborn           & Statistical visualization             & BSD 3-Clause \\
streamlit         & Interactive data exploration UI       & Apache 2.0 \\
kaleido           & Static image export for plotly        & MIT \\
\bottomrule
\end{tabular}%
}
\end{table*}
\label{app:sec_license}
\subsection{Artifacts Licenses}
We provide documentations of the artifact licenses that were used to assist with our data generation and annotation pipeline together with detailed analyses in Table \ref{tab:artifacts_apis}, \ref{tab:artifacts_audio}, \ref{tab:artifacts_packages}.

\subsection{Model Details}
\label{app:model_details}

\paragraph{MOS Predictors}
NISQA, DNSMOS-Pro, UTMOSv2, and Audiobox-Aesthetics were each served via a custom FastAPI inference endpoint on a single NVIDIA Tesla P100-PCIE-12GB GPU paired with an Intel Xeon Gold 6126 CPU @ 2.60 GHz.
 
\paragraph{\audiollm}We evaluate four \audiollm~as automatic judges. Provider, endpoint identifier, input modality, and parameter count are summarized in Table~\ref{tab:audio-llm-judges}. Where the vendor has not published an official parameter count, we report it as \emph{not disclosed}. Table~\ref{tab:audio-llm-judges} summarizes hardware requirements for each LALM judge.
The two Gemini models are accessed via the Google API.
Qwen3-Omni (30B total\,/\,${\approx}$3B active, MoE) requires ${\approx}$60\,GB VRAM at \texttt{bfloat16} and is served on 2$\times$80\,GB GPUs.
Step-Audio-2-mini (8B) requires ${\approx}$16\,GB VRAM but is served on 1$\times$80\,GB GPU to accommodate its custom vLLM backend and audio detokenizer (\texttt{tensor-parallel-size\,=\,1}).
All open-weight models were served via vLLM on NVIDIA A100/H100 GPUs.
Each judge was queried three times per sample per condition; with $\numutterances$ samples $\times$ 4 conditions $\times$ 2 transcript settings, this yields up to 20{,}640 inference calls per model.

\begin{table*}[t]
\centering
\footnotesize
\setlength{\tabcolsep}{3pt}
\caption{Audio-LLM judges and compute requirements. T\,=\,text, I\,=\,image, A\,=\,audio, V\,=\,video. Full model name for row~3: Qwen3-Omni-30B-A3B-Instruct. GPU counts reflect our serving configuration; open-weight models served via vLLM on NVIDIA A100/H100 GPUs.}
\begin{tabular}{@{}cllllrlll@{}}
\toprule
\# & Judge & Provider & Modality & Parameters & VRAM & GPUs & Access \\
\midrule
1 & Gemini 3 Flash     & Google  & T+I+A+V & not disclosed       & -               & -             & API \\
2 & Gemini 3.5 Flash   & Google  & T+I+A+V & not disclosed       & -             & -             & API \\
3 & Qwen3-Omni         & Alibaba & T+I+A+V & 30B/${\approx}$3B (MoE) & ${\approx}$60\,GB & 2$\times$80\,GB & open weights \\
4 & Step-Audio-2-mini  & StepFun & T+A     & 8B                  & ${\approx}$16\,GB & 1$\times$80\,GB & open weights \\
\bottomrule
\end{tabular}

\label{tab:audio-llm-judges}
\end{table*}

\subsection{LLM Prompts}
\label{app:llm_prompts}

Prompts are reproduced exactly as submitted to
\texttt{gpt-5}. Per-utterance slots shown as
placeholders.

\lstdefinestyle{artifactid}{
  basicstyle=\ttfamily\scriptsize,
  breaklines=true,
  breakatwhitespace=false,
  columns=fullflexible,
  keepspaces=true,
  xleftmargin=4pt,
  xrightmargin=4pt,
  frame=none,
  extendedchars=true,
  literate=
    {\textquotedblleft}{{``}}1
    {\textquotedblright}{{''}}1
    {"}{{``}}1
    {"}{{''}}1,
}

\subsubsection*{Phonetic Accuracy}

\begin{tcolorbox}[promptbox, title={\footnotesize Prompt --- Phonetic Accuracy}]
{\footnotesize

\textbf{Task.}
You are a phonetic transcription assistant. Given a sentence and a target
word, produce \textbf{three versions} of the sentence in which the target
word is replaced by three distinct IPA transcriptions in Cartesia
\texttt{sonic-3} phoneme notation, representing decreasing levels of
segmental accuracy (score~1 = most erroneous; score~3 = canonical).

\medskip\hrule\medskip

\textbf{Phoneme Format.}
\label{tab:phoneme-inventory}
Strings are enclosed in \texttt{<{}<}~\ldots~\texttt{>{}>} with
pipe-separated symbols; stress markers precede the stressed syllable.
\textbf{Example:} \textit{Cartesia}~$\to$~\texttt{<{}<k|ɑ|ɹ|ˈ|t|i|ʒ|ə>{}>}

\smallskip
{\tiny
\begin{tabular}{@{}ll@{\quad}ll@{}}
\toprule
\textbf{Ph.} & \textbf{Ex.} & \textbf{Ph.} & \textbf{Ex.} \\
\midrule
\textipa{aI} & \textit{my}     & \textipa{A}  & \textit{cot}     \\
\textipa{aU} & \textit{now}    & \textipa{O}  & \textit{law}     \\
b            & \textit{bed}    & \textipa{OI} & \textit{boy}     \\
d            & \textit{do}     & \textipa{@}  & \textit{sofa}    \\
\textipa{dZ} & \textit{jeans}  & \textipa{E}  & \textit{bed}     \\
\textipa{eI} & \textit{day}    & \textipa{3r} & \textit{bird}    \\
f            & \textit{fish}   & g            & \textit{go}      \\
h            & \textit{hat}    & \textipa{I}  & \textit{bit}     \\
i            & \textit{see}    & \textipa{r}  & \textit{red}     \\
j            & \textit{yes}    & \textipa{S}  & \textit{ship}    \\
k            & \textit{key}    & \textipa{U}  & \textit{book}    \\
l            & \textit{lip}    & \textipa{Z}  & \textit{vision}  \\
m            & \textit{man}    & \textipa{"} & primary stress    \\
n            & \textit{no}     & \textipa{,}  & secondary stress \\
\textipa{oU} & \textit{go}     & \textipa{T}  & \textit{thin}    \\
p            & \textit{pen}    & \textipa{D}  & \textit{this}    \\
s            & \textit{sip}    & \textipa{"\ae} & \textit{cat}   \\
t            & \textit{top}    & \textipa{N}  & \textit{sing}    \\
\textipa{tS} & \textit{church} & u            & \textit{two}     \\
v            & \textit{van}    & w            & \textit{we}      \\
z            & \textit{zoo}    &              &                  \\
\bottomrule
\end{tabular}
}

\medskip\hrule\medskip

\textbf{Input.}\\
\textbf{Sentence:}~\texttt{\{\{INPUT\_UTTERANCE\}\}}\\
\textbf{Target word:}~\texttt{\{\{TARGET\_WORD\}\}}

\medskip\hrule\medskip

\textbf{Output Format.}
Return only valid JSON --- no explanation, markdown, or code fences.
\begin{lstlisting}[style=jsonblock]
{
  "target_word": "<original word>",
  "versions": [
    { "score": 1,
      "error_description": "<substitution + why highly salient>",
      "transcription": "<<IPA-with-error>>",
      "sentence": "<sentence with inline transcription>" },
    { "score": 2,
      "error_description": "<substitution + why moderate>",
      "transcription": "<<IPA-with-error>>",
      "sentence": "<sentence with inline transcription>" },
    { "score": 3,
      "error_description": "Canonical American English pronunciation.",
      "transcription": "<<canonical-IPA>>",
      "sentence": "<sentence with inline transcription>" }
  ]
}
\end{lstlisting}

\medskip\hrule\medskip

\textbf{Constraints.}
\begin{enumerate}
  \item Scores 1 and~2 must not produce a real English word.
  \item Score~1: target the primary stressed syllable; use large
        vowel-class or manner shifts.
  \item Score~2: clearly audible but less dramatic --- consonant
        voicing/manner change, or a one-step vowel shift in a
        secondary-stressed syllable. Do \emph{not} swap schwas in
        fully reduced syllables (inaudible in synthesis).
  \item Do \textbf{not} use word-final /z/$\,\leftrightarrow$\,/s/
        swaps --- the contrast is undetectable in synthesis.
  \item Use only phonemes from the inventory above. Do not introduce
        \texttt{ɚ}, \texttt{ɫ}, \texttt{ɾ}, \texttt{ʔ}, length marks,
        or diacritics absent from the table; write \texttt{ɚ} as
        \texttt{ɝ} (stressed) or \texttt{ə|ɹ} (unstressed).
  \item Score~3 is the unmodified canonical American English
        transcription.
\end{enumerate}

} 
\end{tcolorbox}

\subsubsection*{Lexical Stress}

\begin{tcolorbox}[promptbox, title={\footnotesize Prompt --- Lexical Stress}]
{\footnotesize

\textbf{Task.}
You are a phonetic transcription assistant. Given a sentence and a target
word, produce \textbf{two versions} of the sentence in which the target word
is replaced by a Cartesia \texttt{sonic-3} phoneme string. The two versions
differ only in lexical stress placement.

When stress shifts to a different syllable, apply the phonological changes
that naturally follow in American English:
\textbf{vowel reduction} (newly unstressed full vowels reduce, e.g.\
/ɛ/\,$\to$\,/ə/); \textbf{vowel restoration} (reduced /ə/ in newly stressed
syllables becomes its full counterpart); and any other stress-conditioned
segmental alternations.

\smallskip
Phoneme format and inventory: identical to Phonetic Accuracy above
(see \S\ref{app:ipa_tts}).

\medskip\hrule\medskip

\textbf{Input.}\\
\textbf{Sentence:}~\texttt{\{\{INPUT\_UTTERANCE\}\}}\\
\textbf{Target word:}~\texttt{\{\{TARGET\_WORD\}\}}

\medskip\hrule\medskip

\textbf{Output Format.}
Return only valid JSON --- no explanation, markdown, or code fences.

\begin{lstlisting}[style=jsonblock]
{
  "target_word": "<original word>",
  "versions": [
    { "score": 0,
      "error_description": "<which syllable incorrectly stressed,
        where it should be, and phonological changes applied>",
      "transcription": "<<IPA-with-wrong-stress>>",
      "sentence": "<sentence with inline transcription>" },
    { "score": 1,
      "error_description": "Canonical American English pronunciation
        with correct lexical stress placement.",
      "transcription": "<<canonical-IPA>>",
      "sentence": "<sentence with inline transcription>" }
  ]
}
\end{lstlisting}

\medskip\hrule\medskip

\textbf{Constraints.}
\begin{enumerate}
  \item Target word must be \textbf{polysyllabic}. If monosyllabic,
        return an empty \texttt{versions} array.
  \item Score~0: move primary stress to a \emph{different} syllable.
        If a secondary stress exists, swapping primary and secondary
        is preferred. Apply all resulting phonological changes.
  \item Phoneme changes in score~0 must be a \emph{direct consequence}
        of the stress shift --- no arbitrary substitutions.
  \item Use only phonemes from the inventory (same constraint as
        Phonetic Accuracy).
\end{enumerate}

} 
\end{tcolorbox}

\subsubsection*{Prosodic Stress}

\begin{tcolorbox}[promptbox, title={\footnotesize Prompt --- Prosodic Stress}]
{\footnotesize

\textbf{Task.}
You are a linguistic annotation tool. Given an English sentence, identify
\textbf{function words} (words that should not carry prosodic stress under a
neutral broad-focus reading) and output \textbf{two versions}:
(1)~\textbf{score~1} --- original sentence, unchanged (natural prosody
baseline);
(2)~\textbf{score~0} --- every contiguous run of function words is wrapped
in typographic double quotes \texttt{"~\ldots~"}, forcing a pitch accent on
the quoted span and producing audibly misplaced prosodic stress.

\medskip\hrule\medskip

\textbf{Function-word categories.}
\begin{itemize}
  \item \textit{Determiners:} a, an, the, this, that, these, those
  \item \textit{Prepositions:} in, on, at, to, from, of, for, by, with,
        about, into, through, over, under, between, among, during,
        before, after, against, without
  \item \textit{Auxiliaries:} is, am, are, was, were, be, been, being,
        has, have, had, do, does, did, will, would, shall, should, can,
        could, may, might, must
  \item \textit{Pronouns:} I, me, you, he, him, she, her, it, we, us,
        they, them, my, your, his, its, our, their
  \item \textit{Conjunctions:} and, but, or, nor, so, yet, because,
        although, if, when, while, that, which, who
  \item \textit{Complementizers / relativizers:} that, which, who, whom
  \item \textit{Particles:} to (infinitival)
\end{itemize}

Ambiguous cases: ``that'' as demonstrative~$\to$ function word;
``that'' in contrastive focus~$\to$ content word.
Wh-words in questions~$\to$ content words; as relativizers~$\to$
function words.
Contractions~$\to$ leave unquoted.

\medskip\hrule\medskip

\textbf{Rules.}
\begin{enumerate}
  \item Preserve original spelling, capitalization, and punctuation
        exactly. Sentence-final punctuation stays \emph{outside} any
        quoted span.
  \item A single function word gets its own quote pair; two or more
        adjacent function words share one pair.
  \item Do not add, remove, or reorder words.
\end{enumerate}

\medskip\hrule\medskip

\textbf{Input.}\\
\textbf{Sentence:}~\texttt{\{\{INPUT\_UTTERANCE\}\}}

\medskip\hrule\medskip

\textbf{Output Format.}
Return only valid JSON --- no explanation, markdown, or code fences.
\texttt{target\_word} is always the literal string \texttt{"stress"}
(used for folder naming).

\begin{lstlisting}[style=jsonblock]
{
  "target_word": "stress",
  "versions": [
    { "score": 1,
      "error_description": "Natural prosody -- content words
        stressed, function words reduced.",
      "transcription": "",
      "sentence": "<original sentence unchanged>" },
    { "score": 0,
      "error_description": "Stress placed on \"<region>\"
        (<pos>), \"<region>\" (<pos>).",
      "transcription": "<region1>; <region2>; ...",
      "sentence": "<sentence with function-word runs in \"...\">" }
  ]
}
\end{lstlisting}

\texttt{error\_description} for score~0: begin with \emph{Stress placed
on}, then list each quoted region as \texttt{"<region>"~(<pos>)}, with
multi-word POS joined by `\texttt{+}', e.g.\ \texttt{"on the"
(preposition + determiner)}.

\medskip\hrule\medskip

\textbf{Example.}\\
\textit{Input:} The cat sat on the mat.

\begin{lstlisting}[style=jsonblock]
{
  "target_word": "stress",
  "versions": [
    { "score": 1,
      "error_description": "Natural prosody -- content words stressed,
        function words reduced.",
      "transcription": "",
      "sentence": "The cat sat on the mat." },
    { "score": 0,
      "error_description": "Stress placed on \"The\" (determiner),
        \"on the\" (preposition + determiner).",
      "transcription": "The; on the",
      "sentence": "\u201CThe\u201D cat sat \u201Con the\u201D mat." }
  ]
}
\end{lstlisting}

} 
\end{tcolorbox}

\subsubsection*{Human Plausibility}

\begin{tcolorbox}[promptbox, title={\footnotesize Prompt --- Human Plausibility}]
{\footnotesize

\textbf{Task.}
You are an audio signal processing expert and evaluation dataset designer.
Given a sentence and a target word, produce \textbf{two versions} of the
same utterance at controlled quality levels. Both versions use the same
clean spoken sentence --- the \texttt{sentence} field is identical across
both. The difference is a post-processing \textbf{artifact recipe} applied
after synthesis. Score~1 is an unmodified clean reference; score~0 has
1--2 digital artifacts applied, making it clearly sound artificial or
degraded.

\medskip\hrule\medskip

\textbf{Artifact Types.}
Use these exact identifiers in the \texttt{type} field.

\smallskip
\textit{Compression:}
\texttt{mp3\_compression} (lossy codec smearing, muffled quality);
\texttt{quantization\_noise} (low-bit-depth noise/distortion).

\smallskip
\textit{Clipping \& Distortion:}
\texttt{hard\_clipping} (peaks chopped flat; harsh);
\texttt{soft\_clipping} (warmer harmonic distortion);
\texttt{bit\_crushing} (lo-fi crunch).

\smallskip
\textit{Glitch / Stutter:}
\texttt{buffer\_dropout} (sudden silence or frozen chunk);
\texttt{stutter\_repeat} (rapid looping of a tiny slice);
\texttt{granular\_fragmentation} (micro-segments scrambled).

\smallskip
\textit{Noise:}
\texttt{white\_noise} (broadband injection);
\texttt{pink\_noise} (spectrally weighted);
\texttt{vinyl\_crackle} (crackle and hiss);
\texttt{em\_hum\_50hz} (50\,Hz EM hum);
\texttt{em\_hum\_60hz} (60\,Hz EM hum).

\smallskip
\textit{Time \& Pitch:}
\texttt{time\_stretch\_artifact} (metallic smearing);
\texttt{pitch\_correction\_artifact} (robotic, over-quantized tuning);
\texttt{phase\_cancellation} (comb-filtering from offset copies).

\smallskip
\textit{Digital Errors:}
\texttt{bit\_errors} (random bit flips; sharp pops);
\texttt{sample\_rate\_mismatch} (aliasing or chirping);
\texttt{buffer\_underrun} (choppy, skipping playback).

\smallskip
\textit{Delay-Based:}
\texttt{echo\_feedback} (feedback runaway);
\texttt{resonant\_comb\_filter} (resonant comb);
\texttt{feedback\_squeal} (feedback loop squeal).

\smallskip
\textit{Spectral:}
\texttt{fft\_smearing} (FFT smearing from over-processed EQ);
\texttt{phase\_distortion} (heavy-filter phase distortion);
\texttt{aliasing} (sample-rate-reduction aliasing).

\medskip\hrule\medskip

\textbf{Location Values.}
\texttt{target\_word} : target word's duration only;
\texttt{sentence\_start} : first ${\sim}$20\%;
\texttt{sentence\_middle} : middle ${\sim}$60\%;
\texttt{sentence\_end} : last ${\sim}$20\%;
\texttt{throughout} : full signal.

\medskip\hrule\medskip

\textbf{Severity Values.}
\texttt{mild} : subtle, attentive listening needed;
\texttt{moderate} : clearly audible, intelligibility intact;
\texttt{severe} : highly noticeable, degrades intelligibility.

\medskip\hrule\medskip

\textbf{Input.}\\
\textbf{Sentence:}~\texttt{\{\{INPUT\_UTTERANCE\}\}}\\
\textbf{Target word:}~\texttt{\{\{TARGET\_WORD\}\}}\\
\textbf{Artifact assignment for score~0:}~\texttt{\{\{ARTIFACT\_SELECTION\}\}}

\medskip\hrule\medskip

\textbf{Output Format.}
Return only valid JSON --- no explanation, markdown, or code fences.

\begin{lstlisting}[style=jsonblock]
{
  "target_word": "<original target word>",
  "versions": [
    { "score": 1,
      "sentence": "<original sentence, unchanged>",
      "error_description": "Clean reference -- no artifacts applied.",
      "artifacts": [] },
    { "score": 0,
      "sentence": "<original sentence, unchanged>",
      "error_description": "<listener-perspective description>",
      "artifacts": [
        { "type": "<identifier from list above>",
          "location": "<value from list above>",
          "severity": "<mild | moderate | severe>" }
      ] }
  ]
}
\end{lstlisting}

\medskip\hrule\medskip

\textbf{Constraints.}
\begin{enumerate}
  \item \texttt{sentence} must be \textbf{identical across both versions}.
  \item Score~1 must have \texttt{"artifacts":\ []}.
  \item Score~0 must use \textbf{exactly} the specified artifact type(s) ---
        no substitutions or additions.
  \item At least one score~0 artifact must be \texttt{moderate} or
        \texttt{severe}; mild-only is insufficient.
  \item Use \texttt{location:\ target\_word} for localized glitch artifacts
        (\texttt{buffer\_dropout}, \texttt{stutter\_repeat},
        \texttt{bit\_errors}, \texttt{granular\_fragmentation}).
        Use \texttt{throughout} for diffuse artifacts (noise, compression,
        hum).
  \item Write \texttt{error\_description} from a \emph{listener's}
        perspective, e.g.\ ``A sharp digital stutter interrupts the stressed
        vowel of `eclipse', looping a tiny slice twice before the sentence
        continues.''
  \item Match \texttt{location} and \texttt{severity} to the target word's
        phonetics: stutter/dropout are most salient on vowel-heavy or
        sonorant words; clipping/bit-crushing on fricatives and plosives;
        noise/compression are phoneme-agnostic.
\end{enumerate}

} 
\end{tcolorbox}

\clearpage
\begin{table*}[p]
  \centering
  \rotatebox{90}{%
  \begin{minipage}{0.95\textheight}
  \centering
  \scriptsize
  \renewcommand{\arraystretch}{1.00}
  \setlength{\tabcolsep}{3pt}
  \caption{Per-dimension Kendall's $\tau_b$ against human ground truth. Each data point in the table reports the point estimate and its 95\% percentile bootstrap confidence interval in brackets, computed from 5,000 paired item-level resamples. \textsc{\audiollmsingular} judge rows use the C1 naturalness prompt.}
  \label{tab:tau_bridge_ci_inline}
  \resizebox{\linewidth}{!}{%
  \begin{tabular}{l | ll | llll | llll}
    \toprule
    \textbf{Model} & \multicolumn{2}{c}{\textbf{Word}} & \multicolumn{4}{c}{\textbf{Prosodic}} & \multicolumn{4}{c}{\textbf{Paralinguistic}} \\
    \cmidrule(lr){2-3} \cmidrule(lr){4-7} \cmidrule(lr){8-11}
     & \textbf{Phon.} & \textbf{Lex.} & \textbf{Inton.} & \textbf{Pros.S} & \textbf{Pros.B} & \textbf{Rate} & \textbf{Emo.} & \textbf{Expr.} & \textbf{Spk.} & \textbf{Hum.} \\
    \midrule
    \midrule
    \multicolumn{11}{c}{MOS Models} \\
    \midrule
    AudioBox$_{\mathrm{CE}}$ & -0.051\,{\tiny [-0.202,\,0.104]} & \phantom{-}0.161\,{\tiny [-0.008,\,0.326]} & \phantom{-}\textbf{0.371}{\scriptsize$^{***}$}\,{\tiny \textbf{[0.215,\,0.510]}} & \textbf{-0.241}{\scriptsize$^{**}$}\,{\tiny \textbf{[-0.385,\,-0.086]}} & -0.036\,{\tiny [-0.266,\,0.192]} & \phantom{-}\textbf{0.550}{\scriptsize$^{***}$}\,{\tiny \textbf{[0.462,\,0.623]}} & \phantom{-}0.050\,{\tiny [-0.223,\,0.314]} & \phantom{-}\textbf{0.187}{\scriptsize$^{*}$}\,{\tiny \textbf{[0.022,\,0.348]}} & \textbf{-0.270}{\scriptsize$^{**}$}\,{\tiny \textbf{[-0.456,\,-0.076]}} & \phantom{-}\textbf{0.606}{\scriptsize$^{***}$}\,{\tiny \textbf{[0.484,\,0.696]}} \\
    AudioBox$_{\mathrm{CU}}$ & -0.052\,{\tiny [-0.203,\,0.106]} & \phantom{-}0.015\,{\tiny [-0.155,\,0.180]} & \phantom{-}0.098\,{\tiny [-0.088,\,0.280]} & \phantom{-}0.068\,{\tiny [-0.100,\,0.241]} & \phantom{-}0.048\,{\tiny [-0.179,\,0.271]} & \phantom{-}\textbf{0.485}{\scriptsize$^{***}$}\,{\tiny \textbf{[0.380,\,0.578]}} & -0.025\,{\tiny [-0.268,\,0.236]} & \phantom{-}\textbf{0.418}{\scriptsize$^{***}$}\,{\tiny \textbf{[0.274,\,0.542]}} & -0.177\,{\tiny [-0.364,\,0.024]} & \phantom{-}\textbf{0.511}{\scriptsize$^{***}$}\,{\tiny \textbf{[0.344,\,0.650]}} \\
    AudioBox$_{\mathrm{PC}}$ & -0.058\,{\tiny [-0.209,\,0.098]} & -0.019\,{\tiny [-0.190,\,0.154]} & -0.099\,{\tiny [-0.281,\,0.089]} & \phantom{-}\textbf{0.337}{\scriptsize$^{***}$}\,{\tiny \textbf{[0.196,\,0.473]}} & \phantom{-}0.152\,{\tiny [-0.066,\,0.374]} & \phantom{-}0.020\,{\tiny [-0.124,\,0.162]} & -0.050\,{\tiny [-0.311,\,0.203]} & \phantom{-}\textbf{0.326}{\scriptsize$^{***}$}\,{\tiny \textbf{[0.179,\,0.465]}} & -0.142\,{\tiny [-0.332,\,0.053]} & \textbf{-0.615}{\scriptsize$^{***}$}\,{\tiny \textbf{[-0.692,\,-0.510]}} \\
    AudioBox$_{\mathrm{PQ}}$ & -0.088\,{\tiny [-0.244,\,0.077]} & \phantom{-}0.038\,{\tiny [-0.140,\,0.214]} & \phantom{-}\textbf{0.272}{\scriptsize$^{**}$}\,{\tiny \textbf{[0.101,\,0.430]}} & \textbf{-0.169}{\scriptsize$^{*}$}\,{\tiny \textbf{[-0.325,\,-0.009]}} & -0.079\,{\tiny [-0.298,\,0.154]} & \phantom{-}\textbf{0.447}{\scriptsize$^{***}$}\,{\tiny \textbf{[0.336,\,0.547]}} & \phantom{-}0.015\,{\tiny [-0.235,\,0.270]} & \phantom{-}\textbf{0.304}{\scriptsize$^{***}$}\,{\tiny \textbf{[0.151,\,0.449]}} & \textbf{-0.379}{\scriptsize$^{***}$}\,{\tiny \textbf{[-0.537,\,-0.206]}} & \phantom{-}\textbf{0.606}{\scriptsize$^{***}$}\,{\tiny \textbf{[0.488,\,0.692]}} \\
    DNSMOS-Pro$_{\mathrm{BVCC}}$ & -0.112\,{\tiny [-0.256,\,0.042]} & -0.055\,{\tiny [-0.227,\,0.115]} & \phantom{-}0.092\,{\tiny [-0.093,\,0.264]} & \phantom{-}0.121\,{\tiny [-0.035,\,0.278]} & \phantom{-}0.179\,{\tiny [-0.046,\,0.388]} & \textbf{-0.147}{\scriptsize$^{*}$}\,{\tiny \textbf{[-0.286,\,-0.003]}} & -0.061\,{\tiny [-0.313,\,0.183]} & -0.072\,{\tiny [-0.241,\,0.098]} & -0.196\,{\tiny [-0.382,\,0.003]} & \phantom{-}\textbf{0.450}{\scriptsize$^{***}$}\,{\tiny \textbf{[0.269,\,0.598]}} \\
    DNSMOS-Pro$_{\mathrm{VCC}}$ & \phantom{-}0.005\,{\tiny [-0.147,\,0.158]} & -0.121\,{\tiny [-0.286,\,0.049]} & \phantom{-}0.026\,{\tiny [-0.152,\,0.212]} & -0.048\,{\tiny [-0.203,\,0.115]} & -0.038\,{\tiny [-0.255,\,0.196]} & \phantom{-}0.071\,{\tiny [-0.076,\,0.213]} & \phantom{-}0.128\,{\tiny [-0.117,\,0.357]} & \phantom{-}\textbf{0.406}{\scriptsize$^{***}$}\,{\tiny \textbf{[0.261,\,0.533]}} & -0.092\,{\tiny [-0.284,\,0.109]} & \phantom{-}\textbf{0.397}{\scriptsize$^{***}$}\,{\tiny \textbf{[0.212,\,0.558]}} \\
    NISQA & \phantom{-}0.027\,{\tiny [-0.125,\,0.180]} & -0.047\,{\tiny [-0.220,\,0.124]} & \phantom{-}0.068\,{\tiny [-0.121,\,0.244]} & \phantom{-}0.142\,{\tiny [-0.021,\,0.293]} & \phantom{-}0.099\,{\tiny [-0.126,\,0.322]} & \phantom{-}\textbf{0.236}{\scriptsize$^{**}$}\,{\tiny \textbf{[0.100,\,0.365]}} & -0.015\,{\tiny [-0.266,\,0.227]} & \phantom{-}\textbf{0.347}{\scriptsize$^{***}$}\,{\tiny \textbf{[0.194,\,0.485]}} & -0.036\,{\tiny [-0.237,\,0.164]} & \phantom{-}\textbf{0.377}{\scriptsize$^{***}$}\,{\tiny \textbf{[0.179,\,0.542]}} \\
    UTMOSv2 & \phantom{-}0.000\,{\tiny [-0.155,\,0.155]} & \phantom{-}0.087\,{\tiny [-0.087,\,0.253]} & \phantom{-}\textbf{0.289}{\scriptsize$^{**}$}\,{\tiny \textbf{[0.116,\,0.449]}} & \phantom{-}\textbf{0.200}{\scriptsize$^{*}$}\,{\tiny \textbf{[0.047,\,0.349]}} & -0.024\,{\tiny [-0.256,\,0.202]} & \phantom{-}\textbf{0.390}{\scriptsize$^{***}$}\,{\tiny \textbf{[0.269,\,0.501]}} & -0.062\,{\tiny [-0.313,\,0.193]} & \phantom{-}0.099\,{\tiny [-0.068,\,0.265]} & \phantom{-}0.050\,{\tiny [-0.160,\,0.248]} & \phantom{-}\textbf{0.313}{\scriptsize$^{**}$}\,{\tiny \textbf{[0.098,\,0.510]}} \\
    \midrule
    \multicolumn{11}{c}{AudioLLM Models} \\
    \midrule
    Gemini 3 Flash$^\dag$ & \phantom{-}\textbf{0.323}{\scriptsize$^{***}$}\,{\tiny \textbf{[0.147,\,0.488]}} & \phantom{-}\textbf{0.251}{\scriptsize$^{*}$}\,{\tiny \textbf{[0.055,\,0.441]}} & \phantom{-}0.125\,{\tiny [-0.097,\,0.343]} & \phantom{-}0.125\,{\tiny [-0.069,\,0.314]} & -0.081\,{\tiny [-0.337,\,0.192]} & \phantom{-}0.123\,{\tiny [-0.053,\,0.291]} & \phantom{-}0.089\,{\tiny [-0.215,\,0.381]} & \phantom{-}\textbf{0.274}{\scriptsize$^{**}$}\,{\tiny \textbf{[0.082,\,0.462]}} & \phantom{-}0.036\,{\tiny [-0.207,\,0.285]} & \phantom{-}0.202\,{\tiny [-0.054,\,0.444]} \\
    Gemini 3 Flash$^\triangle$ & -0.022\,{\tiny [-0.207,\,0.164]} & \phantom{-}0.089\,{\tiny [-0.122,\,0.289]} & \textbf{-0.225}{\scriptsize$^{*}$}\,{\tiny \textbf{[-0.439,\,-0.008]}} & \phantom{-}0.099\,{\tiny [-0.099,\,0.281]} & -0.104\,{\tiny [-0.357,\,0.172]} & \phantom{-}0.091\,{\tiny [-0.085,\,0.260]} & \phantom{-}0.000\,{\tiny [-0.284,\,0.297]} & -0.050\,{\tiny [-0.256,\,0.149]} & \phantom{-}0.036\,{\tiny [-0.207,\,0.271]} & \phantom{-}0.121\,{\tiny [-0.145,\,0.372]} \\
    Gemini 3.5 Flash$^\dag$ & \phantom{-}\textbf{0.412}{\scriptsize$^{***}$}\,{\tiny \textbf{[0.256,\,0.553]}} & \phantom{-}\textbf{0.312}{\scriptsize$^{**}$}\,{\tiny \textbf{[0.126,\,0.478]}} & \phantom{-}0.083\,{\tiny [-0.146,\,0.286]} & \phantom{-}\textbf{0.227}{\scriptsize$^{*}$}\,{\tiny \textbf{[0.110,\,0.330]}} & \phantom{-}\textbf{0.276}{\scriptsize$^{*}$}\,{\tiny \textbf{[0.037,\,0.466]}} & \phantom{-}0.116\,{\tiny [-0.057,\,0.271]} & \phantom{-}0.255\,{\tiny [-0.040,\,0.511]} & \phantom{-}0.170\,{\tiny [-0.033,\,0.356]} & \phantom{-}\textbf{0.254}{\scriptsize$^{*}$}\,{\tiny \textbf{[0.020,\,0.447]}} & \phantom{-}0.238\,{\tiny [0.121,\,0.375]} \\
    Gemini 3.5 Flash$^\triangle$ & -0.091\,{\tiny [-0.274,\,0.097]} & \phantom{-}0.082\,{\tiny [-0.123,\,0.282]} & -0.026\,{\tiny [-0.242,\,0.193]} & \phantom{-}0.111\,{\tiny [-0.082,\,0.305]} & \phantom{-}0.230\,{\tiny [-0.033,\,0.484]} & \phantom{-}0.000\,{\tiny [-0.168,\,0.174]} & \phantom{-}0.045\,{\tiny [-0.245,\,0.335]} & \phantom{-}0.135\,{\tiny [-0.068,\,0.330]} & \phantom{-}0.068\,{\tiny [-0.176,\,0.297]} & \phantom{-}0.219\,{\tiny [-0.041,\,0.475]} \\
    Qwen3 Omni$^\dag$ & \phantom{-}0.000\,{\tiny [-0.177,\,0.181]} & -0.124\,{\tiny [-0.295,\,0.089]} & -0.113\,{\tiny [-0.224,\,-0.092]} & \phantom{-}0.039\,{\tiny [-0.158,\,0.222]} & \phantom{-}0.064\,{\tiny [-0.210,\,0.297]} & \phantom{-}\textbf{0.214}{\scriptsize$^{*}$}\,{\tiny \textbf{[0.064,\,0.324]}} & \phantom{-}0.105\,{\tiny [-0.205,\,0.387]} & \phantom{-}0.061\,{\tiny [-0.144,\,0.216]} & \phantom{-}\textbf{0.293}{\scriptsize$^{*}$}\,{\tiny \textbf{[0.055,\,0.508]}} & \phantom{-}0.000\,{\tiny [-0.249,\,0.252]} \\
    Qwen3 Omni$^\triangle$ & \phantom{-}0.173\,{\tiny [-0.019,\,0.359]} & \phantom{-}0.192\,{\tiny [-0.018,\,0.387]} & \phantom{-}0.026\,{\tiny [-0.200,\,0.243]} & -0.103\,{\tiny [-0.300,\,0.091]} & \phantom{-}0.262\,{\tiny [-0.009,\,0.518]} & \phantom{-}\textbf{0.300}{\scriptsize$^{***}$}\,{\tiny \textbf{[0.139,\,0.462]}} & \phantom{-}0.088\,{\tiny [-0.206,\,0.372]} & \phantom{-}0.138\,{\tiny [-0.058,\,0.329]} & \phantom{-}0.228\,{\tiny [-0.015,\,0.455]} & \phantom{-}0.237\,{\tiny [-0.023,\,0.475]} \\
    Step Audio 2 Mini$^\dag$ & -0.018\,{\tiny [-0.203,\,0.174]} & \phantom{-}0.022\,{\tiny [-0.187,\,0.222]} & -0.128\,{\tiny [-0.344,\,0.091]} & -0.080\,{\tiny [-0.275,\,0.119]} & \phantom{-}0.116\,{\tiny [-0.156,\,0.377]} & -0.096\,{\tiny [-0.267,\,0.075]} & -0.132\,{\tiny [-0.422,\,0.159]} & \phantom{-}0.065\,{\tiny [-0.141,\,0.265]} & \phantom{-}0.129\,{\tiny [-0.117,\,0.372]} & \phantom{-}0.073\,{\tiny [-0.190,\,0.332]} \\
    Step Audio 2 Mini$^\triangle$ & \phantom{-}0.019\,{\tiny [-0.171,\,0.206]} & \phantom{-}0.023\,{\tiny [-0.186,\,0.226]} & \phantom{-}0.100\,{\tiny [-0.122,\,0.318]} & \phantom{-}0.040\,{\tiny [-0.162,\,0.238]} & -0.111\,{\tiny [-0.373,\,0.155]} & -0.078\,{\tiny [-0.248,\,0.095]} & -0.039\,{\tiny [-0.333,\,0.250]} & -0.021\,{\tiny [-0.230,\,0.173]} & \phantom{-}0.095\,{\tiny [-0.151,\,0.340]} & \phantom{-}0.036\,{\tiny [-0.234,\,0.305]} \\
    \bottomrule
  \end{tabular}%
  }
  \vspace{4pt}
  \parbox{\linewidth}{\footnotesize
    \textbf{Model Subscripts:} CE\,=\,Content Enjoyment; CU\,=\,Content Usefulness; PC\,=\,Production Complexity; PQ\,=\,Production Quality. DNSMOS-Pro subscripts: BVCC\,=\,trained on BVCC corpus; VCC\,=\,trained on VCC2018 corpus. \textbf{Columns:} Phon.\,=\,Phonetic Accuracy; Lex.\,=\,Lexical Stress; Inton.\,=\,Intonation; Pros.S\,=\,Prosodic Stress; Pros.B\,=\,Prosodic Boundary; Rate\,=\,Speech Rate; Emo.\,=\,Emotional Appropriateness; Expr.\,=\,Expressiveness; Spk.\,=\,Speaker Identity; Hum.\,=\,Human Plausibility. \textbf{Transcript:} $^\dag$\,=\,with transcript; $^\triangle$\,=\,without transcript. \textbf{Bold} = significant ($p<.05$). $^{*}p<.05$, $^{**}p<.01$, $^{***}p<.001$ (two-sided). A dot indicates undefined $\tau_b$ because the model produced a constant prediction.}
  \end{minipage}%
  }
\end{table*}

\clearpage
\begin{table*}[p]
  \centering
  \rotatebox{90}{%
  \begin{minipage}{0.95\textheight}
  \centering
  \scriptsize
  \renewcommand{\arraystretch}{0.95}
  \setlength{\tabcolsep}{3pt}
  \caption{Per-dimension Kendall's $\tau_b$ for \textsc{\audiollmsingular} judges using C2a, C2b, and C3 prompts. Each data point in table reports the point estimate and its 95\% percentile bootstrap confidence interval in brackets, computed from 5,000 paired item-level resamples.}
  \label{tab:tau_lalm_ci_inline}
  \resizebox{\linewidth}{!}{%
  \begin{tabular}{l | ll | llll | llll}
    \toprule
    \textbf{Model} & \multicolumn{2}{c}{\textbf{Word}} & \multicolumn{4}{c}{\textbf{Prosodic}} & \multicolumn{4}{c}{\textbf{Paralinguistic}} \\
    \cmidrule(lr){2-3} \cmidrule(lr){4-7} \cmidrule(lr){8-11}
     & \textbf{Phon.} & \textbf{Lex.} & \textbf{Inton.} & \textbf{Pros.S} & \textbf{Pros.B} & \textbf{Rate} & \textbf{Emo.} & \textbf{Expr.} & \textbf{Spk.} & \textbf{Hum.} \\
    \midrule
    \multicolumn{11}{c}{C2a --- Multi-dimension schema; single holistic score} \\
    \midrule
    Gemini 3 Flash$^\dag$ & \phantom{-}\textbf{0.292}{\scriptsize$^{**}$}\,{\tiny \textbf{[0.144,\,0.413]}} & \phantom{-}0.037\,{\tiny [-0.176,\,0.237]} & \phantom{-}0.197\,{\tiny [0.104,\,0.310]} & \phantom{-}0.141\,{\tiny [0.087,\,0.236]} & \phantom{-}0.064\,{\tiny [-0.211,\,0.316]} & \phantom{-}0.000\,{\tiny [-0.170,\,0.170]} & \phantom{-}0.264\,{\tiny [0.137,\,0.416]} & \phantom{-}0.142\,{\tiny [-0.062,\,0.290]} & \phantom{-}0.177\,{\tiny [0.106,\,0.298]} & \phantom{-}0.135\,{\tiny [0.108,\,0.269]} \\
    Gemini 3 Flash$^\triangle$ & \phantom{-}\textbf{0.206}{\scriptsize$^{*}$}\,{\tiny \textbf{[0.028,\,0.365]}} & \phantom{-}\textbf{0.327}{\scriptsize$^{**}$}\,{\tiny \textbf{[0.162,\,0.468]}} & -0.160\,{\tiny [-0.265,\,-0.097]} & \phantom{-}0.000\,{\tiny [-0.188,\,0.183]} & \phantom{-}\textbf{0.283}{\scriptsize$^{*}$}\,{\tiny \textbf{[0.137,\,0.426]}} & \phantom{-}0.172\,{\tiny [0.011,\,0.288]} & \phantom{-}0.129\,{\tiny [-0.165,\,0.385]} & \phantom{-}\textbf{0.211}{\scriptsize$^{*}$}\,{\tiny \textbf{[0.102,\,0.316]}} & \phantom{-}0.124\,{\tiny [0.100,\,0.246]} & \phantom{-}0.000\,{\tiny [-0.252,\,0.247]} \\
    Gemini 3.5 Flash$^\dag$ & \phantom{-}\textbf{0.486}{\scriptsize$^{***}$}\,{\tiny \textbf{[0.346,\,0.615]}} & \phantom{-}\textbf{0.294}{\scriptsize$^{**}$}\,{\tiny \textbf{[0.100,\,0.485]}} & \phantom{-}0.119\,{\tiny [-0.102,\,0.318]} & \phantom{-}\textbf{0.216}{\scriptsize$^{*}$}\,{\tiny \textbf{[0.029,\,0.381]}} & \phantom{-}0.064\,{\tiny [-0.214,\,0.308]} & \phantom{-}0.000\,{\tiny [-0.172,\,0.172]} & \phantom{-}\textbf{0.396}{\scriptsize$^{**}$}\,{\tiny \textbf{[0.143,\,0.633]}} & -0.031\,{\tiny [-0.230,\,0.171]} & \phantom{-}0.122\,{\tiny [-0.129,\,0.352]} & \phantom{-}0.146\,{\tiny [-0.122,\,0.380]} \\
    Gemini 3.5 Flash$^\triangle$ & \phantom{-}\textbf{0.396}{\scriptsize$^{***}$}\,{\tiny \textbf{[0.245,\,0.536]}} & \phantom{-}\textbf{0.438}{\scriptsize$^{***}$}\,{\tiny \textbf{[0.245,\,0.607]}} & -0.052\,{\tiny [-0.258,\,0.173]} & \phantom{-}0.064\,{\tiny [-0.130,\,0.264]} & -0.104\,{\tiny [-0.356,\,0.168]} & \phantom{-}0.102\,{\tiny [-0.077,\,0.265]} & \phantom{-}0.099\,{\tiny [-0.191,\,0.385]} & \phantom{-}0.170\,{\tiny [-0.029,\,0.354]} & \phantom{-}\textbf{0.296}{\scriptsize$^{*}$}\,{\tiny \textbf{[0.068,\,0.496]}} & \phantom{-}0.115\,{\tiny [-0.153,\,0.352]} \\
    Qwen3 Omni$^\dag$ & \phantom{-}0.097\,{\tiny [-0.099,\,0.232]} & \phantom{-}0.146\,{\tiny [-0.064,\,0.291]} & \phantom{-}0.052\,{\tiny [-0.175,\,0.252]} & -0.101\,{\tiny [-0.245,\,0.100]} & \phantom{-}0.192\,{\tiny [-0.072,\,0.385]} & . & . & \phantom{-}0.182\,{\tiny [0.097,\,0.284]} & \phantom{-}\textbf{0.435}{\scriptsize$^{***}$}\,{\tiny \textbf{[0.228,\,0.623]}} & \phantom{-}0.000\,{\tiny [-0.231,\,0.230]} \\
    Qwen3 Omni$^\triangle$ & \phantom{-}\textbf{0.194}{\scriptsize$^{*}$}\,{\tiny \textbf{[0.094,\,0.288]}} & \phantom{-}\textbf{0.212}{\scriptsize$^{*}$}\,{\tiny \textbf{[0.023,\,0.382]}} & . & -0.141\,{\tiny [-0.237,\,-0.087]} & \phantom{-}0.000\,{\tiny [-0.228,\,0.225]} & . & . & . & -0.124\,{\tiny [-0.246,\,-0.100]} & \phantom{-}0.135\,{\tiny [0.108,\,0.265]} \\
    Step Audio 2 Mini$^\dag$ & \phantom{-}0.000\,{\tiny [-0.184,\,0.187]} & \phantom{-}0.186\,{\tiny [0.099,\,0.292]} & . & -0.101\,{\tiny [-0.245,\,0.101]} & \phantom{-}0.000\,{\tiny [-0.262,\,0.252]} & \phantom{-}0.000\,{\tiny [-0.146,\,0.148]} & \phantom{-}0.213\,{\tiny [0.125,\,0.368]} & -0.147\,{\tiny [-0.242,\,-0.091]} & \phantom{-}\textbf{0.413}{\scriptsize$^{***}$}\,{\tiny \textbf{[0.188,\,0.613]}} & \phantom{-}0.135\,{\tiny [0.104,\,0.275]} \\
    Step Audio 2 Mini$^\triangle$ & \phantom{-}0.041\,{\tiny [-0.149,\,0.227]} & \phantom{-}0.000\,{\tiny [-0.208,\,0.210]} & \phantom{-}0.160\,{\tiny [0.097,\,0.271]} & -0.039\,{\tiny [-0.235,\,0.144]} & -0.111\,{\tiny [-0.374,\,0.151]} & \textbf{-0.221}{\scriptsize$^{*}$}\,{\tiny \textbf{[-0.379,\,-0.049]}} & \phantom{-}0.000\,{\tiny [-0.296,\,0.280]} & -0.165\,{\tiny [-0.357,\,0.031]} & \phantom{-}0.061\,{\tiny [-0.180,\,0.303]} & -0.153\,{\tiny [-0.400,\,0.119]} \\
    \midrule
    \multicolumn{11}{c}{C2b --- Multi-dimension schema; one score per dimension} \\
    \midrule
    Gemini 3 Flash$^\dag$ & \phantom{-}\textbf{0.397}{\scriptsize$^{***}$}\,{\tiny \textbf{[0.290,\,0.504]}} & \phantom{-}0.106\,{\tiny [0.089,\,0.208]} & . & . & \phantom{-}0.196\,{\tiny [0.114,\,0.331]} & . & \phantom{-}0.149\,{\tiny [0.114,\,0.301]} & \phantom{-}0.104\,{\tiny [0.087,\,0.202]} & . & . \\
    Gemini 3 Flash$^\triangle$ & \phantom{-}\textbf{0.194}{\scriptsize$^{*}$}\,{\tiny \textbf{[0.096,\,0.289]}} & \phantom{-}0.186\,{\tiny [0.097,\,0.286]} & . & . & . & . & . & \phantom{-}0.104\,{\tiny [0.087,\,0.202]} & . & . \\
    Gemini 3.5 Flash$^\dag$ & \phantom{-}\textbf{0.514}{\scriptsize$^{***}$}\,{\tiny \textbf{[0.405,\,0.624]}} & \phantom{-}\textbf{0.290}{\scriptsize$^{**}$}\,{\tiny \textbf{[0.166,\,0.401]}} & . & . & \phantom{-}0.137\,{\tiny [0.110,\,0.282]} & \phantom{-}0.126\,{\tiny [0.080,\,0.208]} & \phantom{-}\textbf{0.309}{\scriptsize$^{*}$}\,{\tiny \textbf{[0.149,\,0.462]}} & \phantom{-}\textbf{0.211}{\scriptsize$^{*}$}\,{\tiny \textbf{[0.102,\,0.317]}} & \phantom{-}0.177\,{\tiny [0.106,\,0.298]} & \phantom{-}0.135\,{\tiny [0.108,\,0.268]} \\
    Gemini 3.5 Flash$^\triangle$ & \phantom{-}\textbf{0.210}{\scriptsize$^{*}$}\,{\tiny \textbf{[0.042,\,0.340]}} & \phantom{-}0.106\,{\tiny [0.087,\,0.206]} & \phantom{-}0.113\,{\tiny [0.094,\,0.224]} & . & \phantom{-}0.137\,{\tiny [0.110,\,0.271]} & \phantom{-}0.089\,{\tiny [0.076,\,0.173]} & \phantom{-}0.149\,{\tiny [0.114,\,0.301]} & \phantom{-}0.147\,{\tiny [0.091,\,0.250]} & \phantom{-}0.218\,{\tiny [0.113,\,0.336]} & \phantom{-}0.135\,{\tiny [0.108,\,0.265]} \\
    Qwen3 Omni$^\dag$ & . & . & . & . & . & . & . & . & . & . \\
    Qwen3 Omni$^\triangle$ & . & . & . & . & . & . & . & -0.104\,{\tiny [-0.206,\,-0.087]} & . & . \\
    Step Audio 2 Mini$^\dag$ & . & . & . & . & . & . & . & . & . & . \\
    Step Audio 2 Mini$^\triangle$ & . & . & \phantom{-}0.095\,{\tiny [-0.126,\,0.285]} & . & . & . & \phantom{-}0.149\,{\tiny [0.120,\,0.306]} & . & . & . \\
    \midrule
    \multicolumn{11}{c}{C3 --- Single-dimension focus; one dimension at a time} \\
    \midrule
    Gemini 3 Flash$^\dag$ & \phantom{-}\textbf{0.366}{\scriptsize$^{***}$}\,{\tiny \textbf{[0.258,\,0.469]}} & \phantom{-}0.000\,{\tiny [-0.178,\,0.185]} & \phantom{-}0.083\,{\tiny [-0.142,\,0.286]} & . & \phantom{-}\textbf{0.319}{\scriptsize$^{*}$}\,{\tiny \textbf{[0.166,\,0.465]}} & \phantom{-}0.089\,{\tiny [0.076,\,0.171]} & \phantom{-}0.213\,{\tiny [0.125,\,0.365]} & \phantom{-}0.061\,{\tiny [-0.144,\,0.216]} & . & . \\
    Gemini 3 Flash$^\triangle$ & \phantom{-}\textbf{0.261}{\scriptsize$^{**}$}\,{\tiny \textbf{[0.155,\,0.363]}} & \phantom{-}\textbf{0.267}{\scriptsize$^{*}$}\,{\tiny \textbf{[0.147,\,0.378]}} & \phantom{-}0.000\,{\tiny [-0.218,\,0.216]} & . & \phantom{-}0.055\,{\tiny [-0.210,\,0.308]} & \phantom{-}0.021\,{\tiny [-0.156,\,0.198]} & \phantom{-}0.213\,{\tiny [0.125,\,0.355]} & -0.054\,{\tiny [-0.252,\,0.152]} & . & \phantom{-}0.135\,{\tiny [0.104,\,0.268]} \\
    Gemini 3.5 Flash$^\dag$ & \phantom{-}\textbf{0.355}{\scriptsize$^{***}$}\,{\tiny \textbf{[0.250,\,0.461]}} & \phantom{-}0.178\,{\tiny [-0.019,\,0.326]} & . & . & \phantom{-}0.243\,{\tiny [0.128,\,0.384]} & \phantom{-}\textbf{0.181}{\scriptsize$^{*}$}\,{\tiny \textbf{[0.015,\,0.323]}} & \phantom{-}\textbf{0.459}{\scriptsize$^{**}$}\,{\tiny \textbf{[0.288,\,0.628]}} & -0.060\,{\tiny [-0.256,\,0.144]} & \phantom{-}\textbf{0.286}{\scriptsize$^{*}$}\,{\tiny \textbf{[0.152,\,0.414]}} & \phantom{-}0.135\,{\tiny [0.104,\,0.275]} \\
    Gemini 3.5 Flash$^\triangle$ & \phantom{-}\textbf{0.194}{\scriptsize$^{*}$}\,{\tiny \textbf{[0.094,\,0.286]}} & \phantom{-}0.170\,{\tiny [-0.039,\,0.351]} & -0.155\,{\tiny [-0.318,\,0.071]} & \phantom{-}0.100\,{\tiny [0.085,\,0.198]} & \phantom{-}\textbf{0.276}{\scriptsize$^{*}$}\,{\tiny \textbf{[0.029,\,0.465]}} & \phantom{-}\textbf{0.232}{\scriptsize$^{**}$}\,{\tiny \textbf{[0.062,\,0.381]}} & \phantom{-}\textbf{0.303}{\scriptsize$^{*}$}\,{\tiny \textbf{[0.041,\,0.515]}} & \phantom{-}0.043\,{\tiny [-0.165,\,0.245]} & \phantom{-}0.218\,{\tiny [0.113,\,0.344]} & -0.135\,{\tiny [-0.275,\,-0.108]} \\
    Qwen3 Omni$^\dag$ & . & . & . & . & \phantom{-}0.137\,{\tiny [0.110,\,0.282]} & \textbf{-0.184}{\scriptsize$^{*}$}\,{\tiny \textbf{[-0.349,\,-0.013]}} & . & \phantom{-}0.026\,{\tiny [-0.185,\,0.225]} & . & \phantom{-}0.165\,{\tiny [-0.106,\,0.416]} \\
    Qwen3 Omni$^\triangle$ & \phantom{-}0.136\,{\tiny [0.084,\,0.226]} & \phantom{-}\textbf{0.243}{\scriptsize$^{*}$}\,{\tiny \textbf{[0.119,\,0.354]}} & . & . & \phantom{-}\textbf{0.298}{\scriptsize$^{*}$}\,{\tiny \textbf{[0.034,\,0.552]}} & \phantom{-}0.050\,{\tiny [-0.127,\,0.224]} & -0.149\,{\tiny [-0.292,\,-0.114]} & \textbf{-0.214}{\scriptsize$^{*}$}\,{\tiny \textbf{[-0.406,\,-0.015]}} & . & \phantom{-}0.139\,{\tiny [-0.135,\,0.345]} \\
    Step Audio 2 Mini$^\dag$ & . & . & . & . & . & . & . & \phantom{-}0.104\,{\tiny [0.087,\,0.202]} & . & \phantom{-}0.000\,{\tiny [-0.269,\,0.267]} \\
    Step Audio 2 Mini$^\triangle$ & . & -0.108\,{\tiny [-0.215,\,-0.090]} & . & \phantom{-}0.100\,{\tiny [0.085,\,0.198]} & . & . & . & \phantom{-}0.069\,{\tiny [-0.130,\,0.265]} & . & \phantom{-}0.038\,{\tiny [-0.220,\,0.296]} \\
    \bottomrule
  \end{tabular}%
  }
  \vspace{4pt}
  \parbox{\linewidth}{\footnotesize
    \textbf{Columns:} Phon.\,=\,Phonetic Accuracy; Lex.\,=\,Lexical Stress; Inton.\,=\,Intonation; Pros.S\,=\,Prosodic Stress; Pros.B\,=\,Prosodic Boundary; Rate\,=\,Speech Rate; Emo.\,=\,Emotional Appropriateness; Expr.\,=\,Expressiveness; Spk.\,=\,Speaker Identity; Hum.\,=\,Human Plausibility. \textbf{Transcript:} $^\dag$\,=\,with transcript; $^\triangle$\,=\,without transcript. \textbf{Bold} = significant ($p<.05$). $^{*}p<.05$, $^{**}p<.01$, $^{***}p<.001$ (two-sided). A dot indicates undefined $\tau_b$ because the model produced a constant prediction; no confidence interval is reported for undefined cells.}
  \end{minipage}%
  }
\end{table*}


\end{document}